\documentclass[aps,a4paper,onecolumn,showpacs,showkeys,preprintnumbers,amsmath,amssymb,nofootinbib,10pt]{revtex4}
\usepackage{url,hyperref,lineno,microtype}
\usepackage{color}
\usepackage{graphicx}
\usepackage{amsmath}
\begin{document}
\title{Trajectories entropy in dynamical graphs with memory} 
% Affiliations should be keyed to the author's name with superscript numbers and be listed as follows: Laboratory, Institute, Department, Organization, City, State abbreviation (USA, Canada, Australia), and Country (without detailed address information such as city zip codes or street names).
% If one of the authors has a change of address, list the new address below the correspondence details using a superscript symbol and use the same symbol to indicate the author in the author list.
\author{Francesco Caravelli}
\affiliation{Invenia Labs, 27 Parkside Place, Parkside, Cambridge  CB1 1HQ, UK and\\
London Institute of Mathematical Sciences, 35a South Street, London W1K 2XF, UK and \\ 
Department of Computer Science, University College London, Gower Street, London WC1E 6BT, UK  }
% The Corresponding Author should be marked with an asterisk
% Provide the exact contact address (this time including street name and city zip code) and email of the corresponding author
%\def\corrAuthor{Francesco Caravelli}
%\def\corrAddress{Invenia Labs}
%\def\corrEmail{francesco.caravelli@invenialabs.co.uk}

%\onecolumn
%\firstpage{1}

%\title[Graph complexity]{Complexity of dynamical graphs with memory} 

%\author[\firstAuthorLast ]{\Authors} %This field will be automatically populated
%\address{} %This field will be automatically populated
%\correspondance{} %This field will be automatically populated

%\extraAuth{}% If there are more than 1 corresponding author, comment this line and uncomment the next one.
%\extraAuth{corresponding Author2 \\ Laboratory X2, Institute X2, Department X2, Organization X2, Street X2, City X2 , State XX2 (only USA, Canada and Australia), Zip Code2, X2 Country X2, email2@uni2.edu}
\begin{abstract}

%%% Leave the Abstract empty if your article falls under any of the following categories: Editorial Book Review, Commentary, Field Grand Challenge, Opinion or specialty Grand Challenge.

In this paper we investigate the application of non-local graph entropy to evolving and dynamical graphs. The measure is based upon the notion of Markov diffusion on a graph, and relies on the entropy applied to trajectories originating at a specific node. In particular, we study the model of reinforcement-decay graph dynamics, which leads to scale free graphs. We find that the node entropy characterizes the structure of the network in the two parameter phase-space describing the dynamical evolution of the weighted graph. We then apply an adapted version of the entropy measure to purely memristive circuits. We provide evidence that meanwhile in the case of DC voltage the entropy based on the forward probability is enough to characterize the graph properties, in the case of AC voltage generators one needs to consider both forward and backward based transition probabilities. We provide also evidence that the entropy highlights the self-organizing properties of memristive circuits, which re-organizes itself to satisfy the symmetries of the underlying graph.\footnote{Part of this work was presented at the 7th workshop on Guided Self-Organization, Freiburg, Germany (2014) and at the Symposium on Complexity, Computation and Criticality in Sydney (2015).}

%\tiny
 %\keyFont{ \section{Keywords: graphs with memory, graph complexity, scale free graphs, memristors}\\ } %All article types: you may provide up to 8 keywords; at least 5 are mandatory.
\end{abstract}


\maketitle

%%%%%%%%%%%%%%%%%%%%%%%%%%%%%%%%%%%%%%%%%%%%%%%%%%%%%%%%%%%%%%%%%%%%%%%%%%%%%%%%%%%%%%%%%%%%%%%%%%%%%%%%%%%%%%%%%%%%%%%%%%%%%%%%%%%%%%%%%%%%%%%%%%%%%%%%%%%%%%%%%%%%%%%%%%%%%%%%%%%%%%%%%%%%%%%%%%%%%%%%%%%%%%%%%%%%%%%%%%%%%%%%%%%%%%%
%%% The sections below are for reference only.
%%%
%%% For Original Research Articles, Clinical Trial Articles, and Technology Reports the section headings should be those appropriate for your field and the research itself. It is recommended to organize your manuscript in the
%%% following sections or their equivalents for your field:
%%% Abstract, Introduction, Material and Methods, Results, and Discussion.
%%% Please note that the Material and Methods section can be placed in any of the following ways: before Results, before Discussion or after Discussion.
%%%
%%%For information about Clinical Trial Registration, please go to http://www.frontiersin.org/about/AuthorGuidelines#ClinicalTrialRegistration
%%%
%%% For Clinical Case Studies the following sections are mandatory: Abstract, Introduction, Background, Discussion, and Concluding Remarks.
%%%
%%% For all other article types there are no mandatory sections.
%%%%%%%%%%%%%%%%%%%%%%%%%%%%%%%%%%%%%%%%%%%%%%%%%%%%%%%%%%%%%%%%%%%%%%%%%%%%%%%%%%%%%%%%%%%%%%%%%%%%%%%%%%%%%%%%%%%%%%%%%%%%%%%%%%%%%%%%%%%%%%%%%%%%%%%%%%%%%%%%%%%%%%%%%%%%%%%%%%%%%%%%%%%%%%%%%%%%%%%%%%%%%%%%%%%%%%%%%%%%%%%%%%%%%%%

\section{Introduction}

The theory of complex networks has found applications in many fields across the natural sciences.
Until recently, most of the studies performed in complex networks concerned static graphs.   Preferential attachment \cite{barev} is the most well known mechanism for constructing scale-free networks, and it concerns in fact with a specific type of graphs growth  in order to explain the properties of realistic, well known networks (rich gets richer phenomenon). Scale free networks are defined by  the distribution $P(k)$ of the degrees of connectivity, which obeys a power law  $P(k\gg1)\approx k^{-\rho}$, with
$k$ being the number of connections of a given node. The exponent $\rho$ is typically in the range between 2 and 3  ; moreover, many realistic networks turn out being also small-world  \cite{ws, barev,caldarelli,dyncn,newman,estrada}.

Several models in which preferential attachment is an emergent property have been proposed in the physics literature \cite{fractalweb,es,sk,kleinberg,gfwn,cc,ikeda}, and inspired both by ants and memristors, we proposed a model of dynamical graphs in which two competing mechanisms take place: a graph ``evaporation" (or decay), and a reinforcement process due to the diffusion of particles on the graph.  
In this model, memory is represented by non-markovianity, as the reinforcement process changes the particles' hopping probabilities.

Recently, the interest has shifted towards the understanding of dynamical phenomena \textit{of} graphs. In this article, we present new results on a class of dynamical models that gives rise to scale-free graphs by means of what we call {\it memory}. 
Memory is an ubiquitous and necessary requirement to give rise to complex phenomena in plenty of contexts. 
The main result obtained in \cite{our} is that the interplay between random growth of the graph, decay of the links and their strengthening performed by random walkers hopping over them, leads to the generation of scale free graphs. Thus, although the model is local (in the sense of evolution rules which are local on the graph), it is from the temporal point of view a non-local one. Interestingly, real condensed-matter systems show some degree of memory in their response functions (e.g., its resistance) when subject to external perturbations \cite{memsys}. Similarly, a memory mechanism is used as an optimization procedure by ants in order to find the shortest path, by reinforcing with pheromones the most walked paths \cite{ants, ants2},  and has been shown to be employed by networks of memristors (resistors with memory) \cite{williams,chua} to solve optimization problems such
as the maze \cite{paralleldv} or other  shortest-path problems \cite{shortpdv}. Recently, it has been argued that several ``hard" problems can be solved in polynomial time using memristors \cite{npmem}.

In the case of dynamical graphs it is hard (if not impossible) to characterize with only a few parameters the  topological properties of the dynamics, in particular if the rules of evolution are non-local, as for instance in the case of memristors. Following this line of thought, we tentatively study the properties of the resulting graph using a measure of graph entropy introduced in \cite{rankingcar}. This measure of graph entropy is based on the idea that for the case of graphs, in order to give a non-local characterization of a node based on the macroscopic connectivity properties, one needs to consider higher order loops in the network. 
In the present paper we provide evidence that for the specific case of two specific dynamical graph models, such as the reinforcement-decay and memristive circuit models, the evolution properties can be to some extent characterized by a non-local graph entropy.

The use of non-local information theoretic measures is inspired by the macroscopic notions of entropy introduced in \cite{Lloyd,Lindgren} and applied to the Markov chain transition probabilities based on dynamical graphs. 
If $M$ denotes the Markov transition matrix, one can define the (local) entropy of a state $i$ as:
\begin{equation}
S(i)=-\sum_j M_{ij} \log(M_{ij}),
\label{eq:ent}
\end{equation}
with $M_{ij}$ being the Markov operator. The transition matrix $M$ can be derived from a graph $\mathcal G$, given an elementwise non-negative adjacency matrix $A$,  by normalizing over rows or columns. From the point of view of graph theory, eqn. (\ref{eq:ent}) is completely local. In order to account for non-local effects, such entropy necessitates an extension to account non-local (global) effects, such as loops. Similar ideas were also considered in the information theory community \cite{Ekroot}, and in the complex networks community \cite{severini,bianconient}. In the present paper we consider the extension considered in \cite{rankingcar}, which relies on the definition of Markov probabilities not on the local diffusion probability at a node, but on the probability of trajectories \textit{originated} at a specific node.

The paper is organized as follows. In Section \ref{sec:modelrd}we provide the algorithm for the toy model of reinforcement-decay introduced in a previous work, and give further arguments for its robustness. In Section \ref{sec:memr} 
we introduce linear memristors and their dynamical behavior. In Section \ref{sec:gent} we recall the graph entropy measure later used to analyze the properties of dynamical graphs. We then apply the graph entropy in Sections \ref{sec:rdcomp} and \ref{sec:memrcomp} for the reinforcement-decay model and memristive circuits respectively. Conclusions follow.
%\begin{methods}

\section{Models}
\subsection{Reinforcement-decay dynamical graphs} \label{sec:modelrd}
\subsubsection{Scale free graphs from memory decay}

In this section we recall the graph evolution model introduced in \cite{our} in order to study a simplified dynamics mimicking systems of ants (reinforcing walkers) and as a toy model for studying graphs with memory. 

{\em Model.---} The algorithm to dynamically modify the graph is based upon the following steps:\\
%\begin{enumerate}[(i)]

{\em a) Initialization:} We start with a weighted random graph of $N_0$ nodes. The weights are drawn with constant probability in $[0,1]$, and $P \leq N_0$ particles are randomly placed. 

After initialization, a cycle of the algorithm consists of the steps of Hopping, Strengthening/Decay, and Growth:\\

{\em b) Hopping:} We let the particles hop between nodes $i$ and $j$ with probability $p_{ij}$ proportional to the link strength $ p_{ij}={A_{ij}}/{\sum_j A_{ij}}$, where $A_{ij}$ is the weighted adjacency matrix of the graph.\\

{\em c) Strengthening/Decay:} All the links hopped on by the particles in the last  $L$ steps are reinforced by $\gamma$. Links with strength less than threshold $L_d$ decrease their strength by $\alpha$, with probability $p_d$, and are removed when they reach a negative weight.\\

%A graphical representations of steps (b) and (c) are shown in Fig. \ref{fig:walkdecay}. 

A critical ingredient to obtain scale free graphs was shown to be:

{\em d) Growth:} At this step, a new node is added (and with probability $p_p$ a new particle is placed on it). The new node connects to each of the previously existing nodes with probability $p_{nl}$, here chosen to be one, and with random weights with constant probability in $[0,1]$. %\textcolor{red}{R1: In d) the parameter $p_nl$ is introduced. Afterwards we realize that the nl isn't two indexes but it stands for "new link". Please use that wording near the first apparition of the parameter. This is comment is valid for all named parameters. What are they? What do they model?}

%In this section we introduce a random mechanism for the preservation of the global properties of the resulting graph. We call this process random knocking.

%{\em Model.---} The algorithm to create and update the network consists of the following four steps:\\
%\begin{enumerate}[(i)]
%{\em d) Random Knocking:} Random Knocking: at each time step, choose a node $i$, and set the links between every node $i$ and another node $j$ to a random value $w_{ij}$.

%\textcolor{red}{explain here better}

The memory feature of this model lays in the non-markovianity of the particles hopping. Each time a particle hops on an existing link this is reinforced, and thus the probability of hopping on it afterwards increases. We note that the decay process competes with the reinforcing one. The interplay between these two phenomena has been shown to lead to a critical growth of the effective graph, obtaining asymptotic degree distributions which are scale free. The observation that the growth of the graph is a crucial step for scale free degree distribution  is made by imposing the growth to stop at a certain maximum number of nodes $N_c$. After the growth stops, it is observed that graph properties change abruptly and the scale free degree distribution is lost. 
The reinforcement-decay graph evolution is a rich-gets-richer mechanism, and was inspired by travelling ants leaving an evaporating track of pheromones; also memristors have similar dynamical properties.  As described in the Appendix of \cite{our}, the observation of the scale free degree distribution law can be made precise in the statistical sense and derived analytically. Here we provide the background, giving further arguments of why the algorithm is robust. 

It is easy to argue that the degree  $k_s$ of the node $s$ has an effective evolution equation of the form:
\begin{equation}
k_s(t+1)=k_s(t)- \alpha k_s(t) + p_s \sum_{i=1}^N w_{s i},
\end{equation}
where $w_{s i}$ are independent random variables drawn from a constant probability distribution $P(x)=1$, and $p_s=2 \rho/N$ to match the evolution obtained in \cite{our}. We note that for $N\gg 1$, due to the central limit theorem, we have that $\sum_{i=1}^N w_{s i}\approx \mathcal N(\frac{N}{2},  \frac{\sqrt{N}}{\sqrt{12}})$, where $1/12$ is the variance of the distribution, and $\frac{N}{2}$ is its mean.
This implies that in the large $N$ limit, we can describe the evolution of the degree of the graph as a stochastic differential equation of the form:
\begin{equation}
dk_s(t)=\left(p_s \frac{N}{2}- \alpha k_s(t)\right) dt + p_s \tilde \sigma  dW_t,
\label{eq:stocdiff}
\end{equation}
where $dW_t$ is an \textit{effective} Wiener differential, and where we defined $\tilde \sigma =\frac{\sqrt{N}}{\sqrt{12}}$, and introduced $dk_s(t)=k_s(t+1)-k_s(t)$. Using standard formulae for the mean and the variance, we can thus write the compact effective equation, and replacing the right scaling $p_s\rightarrow 2\rho/N$ to account for the linear growth of the degree:
\begin{eqnarray}
dk_s(t) &=&\left(\rho- \alpha k_s(t)\right) dt + \frac{2 \rho}{N} \tilde \sigma  dW_t \nonumber \\
&=&\left(\rho- \alpha k_s(t)\right) dt + \frac{\rho}{\sqrt{3 N} }  dW_t
\label{eq:stocdiff2}
\end{eqnarray}
The stochastic differential equation of eqn. (\ref{eq:stocdiff2} ) of the same form as the one obtained in \cite{our}, but with the addition of a stochastic term. We note that eqn. (\ref{eq:stocdiff2}) is a stochastic differential equation of the form:
\begin{equation}
dk = (a k + c) dt + (b k+d) dW_t,
\label{eqn:easyform}
\end{equation}
for the function $k(t)$, where the constants are given by
\begin{eqnarray}
a&=&-\alpha\\
b&=& 0\\
c&=& \rho \\
d&=&\frac{\rho}{\sqrt{3 N} } .
\end{eqnarray}
If we define the function $\Phi_t$ given by,
\begin{equation}
\Phi_t (t,W_t)\equiv  \exp\left( (a-b^2/2)t + b W_t \right) = \exp\left( -\alpha t \right),
\end{equation}
the solution of the differential equation of eqn. (\ref{eq:stocdiff2}) is given by \cite{kloedenplaten}:
\begin{eqnarray}
k_s(t)&=& \Phi_t \left(k^0_s+(c-bd) \int_0^t \Phi_s^{-1} ds+d \int_0^t \Phi_s^{-1}dW_s\right) \nonumber \\
&=&e^{ -\alpha  t }\left(k^0_s+\rho \int_0^t e^{ \alpha s } ds+\frac{\rho}{\sqrt{3 N} } \int_0^t e^{ \alpha  s} dW_s\right)\nonumber \\
&=& e^{ -\alpha t }\left(k^0_s+\rho \frac{e^{ \alpha  t }-1}{\alpha }+\frac{\rho}{\sqrt{3 N} } \int_0^t e^{ \alpha s}dW_s\right)\nonumber\\
\label{eq:impactlog}
\end{eqnarray}
where $k^0_s$ represents the initial condition. Using the fact that $\langle \int^t I(s) dW_s \rangle_{W_t} =0$ for all smooth and deterministic processes $I(s)$, we obtain the solution as a function of $t$:
\begin{equation}
\langle k_s(t) \rangle_{W_t} = e^{ -\alpha t }\left(k^0_s+\rho \frac{e^{ \alpha  t }-1}{\alpha }\right).
\end{equation}
The constant $k^0_s$ can be related to the node $s$ using the boundary conditions at a certain initial time $t_0$.  It is easy to evaluate the variance of this process at time $t$, being given by the stochastic term. Using the It\^{o} isometry formula, we note that $\text{Var}[k_s(t)]=\frac{\rho^2}{3N}\int_0^t e^{2 \alpha s} ds=\frac{\rho^2}{3 N}\frac{(e^{2 \alpha t}-1)}{2 \alpha } $. This solution is the same as the one obtained in \cite{our}, which was confirmed by means of numerical simulations, and exhibiting very robust scale free distributions under parameter perturbations.  The robustness can be explained observing that the variance of the process scales as $1/N$. This implies that the larger the graph, the smaller the deviation from the asymptotic distribution.

\subsection{Memristive circuits}\label{sec:memr}

A memristor can be thought simply as a dynamical resistance which depends on an internal state variable.
The internal variable, the ``memory", satisfies a dynamical law which depends either on the applied voltage or on the current.
Memristors are passive components which perform analog computation:
computing in memory is in fact one of the newly formed paradigms in which novel
non-Turing computational scenarios involve memory elements which store and process information 
in the same physical location \cite{naturemax}. Although this new form of computation could, in principle, be 
performed using CMOS technology, the most promising and energy-effective implementation 
relies upon the use of memristive, memcapacitive and meminductive units \cite{memel}. 
These components can find a plethora of applications in electronics for building
bio-inspired circuitry aimed at performing neuromorphic computation. 

One of the interesting features of memristors is that these can be used both in analog or digital mode, and in combination with CMOS components \cite{memsys}. Several applications have been found for memristors, such as their use
in memcapacitive neural networks, and the possibility of performing logical operations directly within memory using memelements.
This latter feature is of particular interest to solve the long-standing von Neumann bottleneck problem of modern computers and solve NP-hard problems in polynomial \textit{time} \cite{npmem}.
Being these passive component, the use of memelement for low energy dissipating circuits is a further appealing feature.

Memristive components are involved in the solution of optimization problems. It is thus of paramount importance to know how fast circuits made of memristor reach an asymptotic equilibrium configuration of the memristances. This implies that studying the emergence of the solution of an optimization problem can be thought as a relaxation phenomenon. This problem thus become a typical thermalization problem in non-equilibrium statistical physics. In this paper we will try to infer global, non-local properties of the dynamics of memristive circuits using insights which arose in the context of network theory. 

A memristive component can be described by the following set of equations
\begin{eqnarray}
V(t) &=& R(w,t) I(t), \\
\dot w &=& f (w, I) ,
\end{eqnarray}
where $I(t)$ is the current in the memristor at time t, $V(t)$ is the applied voltage, $R$ is memristance
which depends on the state of the system and can vary in time, $w$ is a set of $n$ state variables describing
the internal state of the system, and $f$ is a continuous $n$-dimensional vector function.

For the present paper we consider the case of a linear memristor of the HP-type \cite{williams}, which can be described by the equations:
\begin{eqnarray}
V(t) &=& R(w,t) I(t) \\
\dot w &=& p \frac{\mu R_{on} }{2 d^2} I = \frac{p}{\beta} R_{on} I \\
R(w,t)&=&R_{on} (1-w(t))+R_{off} w(t) \nonumber \\
&=& R_{on}[1+(r-1) w(t)],
\end{eqnarray}
where $p=\pm1$ and represents the polarity of the memristor, $R_{on}$ is the limiting resistance when the memristor is in the conducting phase, $R_{off}$ is the resistance in the insulating phase ; we defined $r=\frac{R_{off}}{R_{on}}$, usually assumed to satisfy the relation $r\gg 1$.\footnote{We use here the convention where $R_{on}$ corresponds to $w=0$ and $R_{off}$ to $w=1$. Thus the memristors considered here have switched polarity with respect to those studied by \cite{williams}. This model is dynamically not equivalent to the one in \cite{williams} because of the switched polarity. However, the two models are related by the linear transformation $w^\prime=1-w$, which preserves the position of the boundary values of the internal parameters.}

This is the simplest model proposed in the literature, but more realistic ones include state decay. In fact, \cite{Ohno2011,Avizienis2012,Carbajal2015} realistic physical models include also an extra term proportional to the weight.  The simplest way to include such decay is to extend the dynamics for internal parameter dynamics equation with a linear term which produces an exponential decay. The extended equation takes the form:
\begin{equation}
\dot w=\frac{p}{\beta}R_{on} I+\alpha w.
\end{equation}
where $\alpha$ represents the decay parameter. Assuming for instance that $I$ is constant asymptotically, dynamically one obtains a fixed point in the internal memory parameter, $w(\infty)\propto I$, which can be different from the boundary values $w=0,1$. If $\alpha>0$, in the absence of external current the memristor decays to obtain the boundary resistance value $R_{off}$.

A network of pure memristors satisfies the laws of standard, linear circuit theory. Given in fact a generic network of resistances, the circuit voltages and currents satisfy the constraints given by:
\begin{eqnarray}
\vec v&=&\hat R\left(w(t)\right) \vec i+\vec s \\
\hat G \vec i&=&0
\label{eq:circuit}
\end{eqnarray}
where the first equation represents simply the relation between the voltage on each branch $k$ of the network $v_k$, the current in that branch $i_k$, the eventual voltage sources $s_k$ and each resistance. The matrix $\hat G$ simply implements the conservation of currents at each node, which can be thought as a constraint.
For the case of memristors, the matrix $\hat R$, and each diagonal element contains the resistance of each memristor at time $t$, $R_{kk}(t)\equiv R_k\left(w_k(t)\right)$. If the memristors are of the HP type, then one has to introduce also an equation for the internal memory parameters $\dot w_k(t)=f(w_k,I_k)$, and impose that constraints on their domain, e.g. $0\leq w_k(t)\leq1$.

In this paper we study the specific network configuration, made of $L$ layers and which can be defined recursively. Each layer $k$ contains $2^k$ memristors connected in triangles, representing thus a tree. The node at the top of the network is connected to one side of the voltage generator, and those at the bottom are connected to the other one, and all connected to the ground. We consider this specific graph due to its many symmetries (i.e. intuitively we expect that nodes on each layer should have the same properties and current configurations for instance) and for the sake of simplicity, since its implementation can be easily made recursive. \footnote{The author is happy to disclose the code made to generate these simulations upon request.}

\begin{figure}
\begin{center}
\includegraphics[width=8cm]{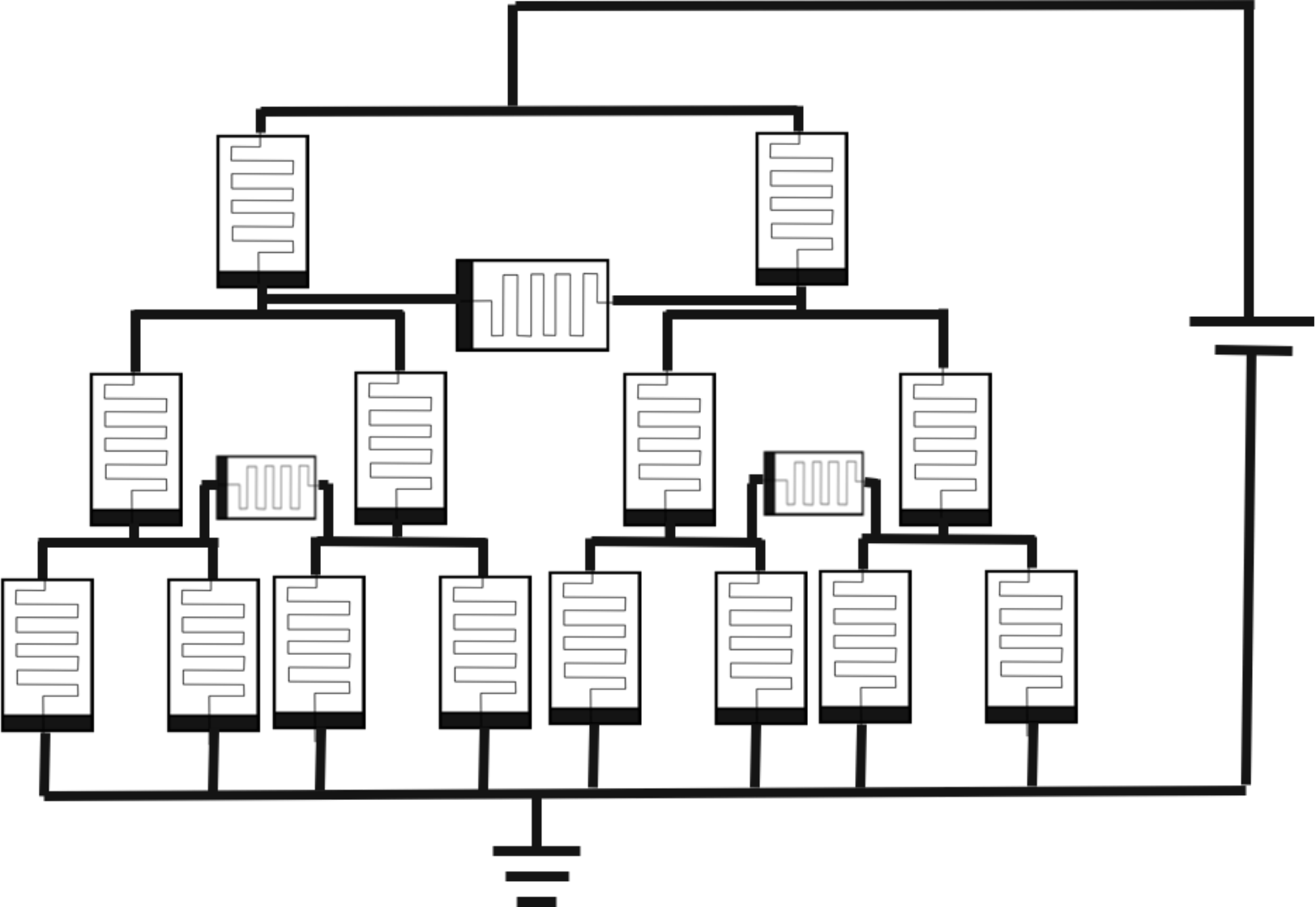}
\end{center}
\caption{Memristive circuit defined recursive as on a tree and made of $L$ layers. The nodes in the bottom layer are connected to the ground.}
\label{fig:TREE}
\end{figure}

\subsection{Graph entropy}\label{sec:gent}
We recall the network entropy measure introduced in \cite{rankingcar}. The goal is to obtain a graph entropy, based on the Markov transition probability on a graph, which is able to characterize nodes individually.

Given a graph represented by an elementwise non-negative adjacency matrix $A$, the Markov forward (backward) transition probabilities can be derived using $M_{ij}=\frac{A_{ij}}{\sum A_{ij}}$ ($M_{ij}=\frac{A_{ij}}{\sum_j A_{ji}}$). As we will see shortly, the entropy is based upon the notion of probability applied to a particular trajectory in the set of states of a Markovian discrete dynamics and originating at a particular state. 
Given a the Markovian transition probabilities between two states (nodes of the graph), it is possible to derive the probability of certain trajectories $\gamma$, here denoted as $\tilde M(\gamma)$. We denote $\{\gamma_i^k\}$ the set of trajectories of length $k$ and originating at $i$.
For instance, if the path is given by $\gamma=\{i,j,k\}$, the probability of that trajectory will be given by $M(\gamma)=M_{ij} M_{jk}$. Since $\sum_{\{\gamma_i^k\}}  \tilde M(\gamma_i^k) =\sum_{j_1,j_2,\cdots,j_k} \tilde M(\{i,j_1,\cdots,j_k \})=1\ \forall\ k$, it is easy to see that $(\{\gamma_i^k\},\tilde M(\{\gamma_i^k\})$ is a probability space.  We note that $\tilde M(\gamma_i^k)$ is a node-dependent quantity, is non-local, and can be interpreted as a probability measure over the space of walk of fixed length originated at a node. 

Since the goal is to introduce an entropy over such probability space, we note that the first possibility for such an entropy is given by:
\begin{equation}
(\ ^{k+1} \vec S)_i =-\frac{1}{k}\sum_{\{\gamma^k_i\}} \tilde M(\gamma) \log \left( \tilde M(\gamma)  \right);
\label{eq:entropy}
\end{equation}
however, due to the Ces\'{a}ro mean rule, in the limit $k\rightarrow \infty$ the entropy becomes independent of the initial node $i$. Although this quantity, while being important, does not provide a measure which differentiates the nodes. An alternative normalization has thus to be considered.

As shown in \cite{rankingcar}, interesting properties arise if one chooses a different and multiplicative normalization. We introduce the following normalization: 
\begin{equation}
\ ^{k+1} \vec S_\epsilon =-\epsilon^{k-1} \sum_{\{\gamma^k_i\}} \tilde M(\gamma) \log \left( \tilde M(\gamma)  \right),
\label{eq:entropy}
\end{equation}
where $\tilde M(\gamma)$ is the probability associated with the trajectory $\gamma$. The parameter $\epsilon$ can be chosen arbitrarily, but in order to guarantee convergence in the limit $k\rightarrow \infty$, it is necessary to impose $\epsilon<1$. We note that evaluating the entropy from eqn. (\ref{eq:entropy}) is computationally hard, as it involves the evaluation of an exponential number of trajectories of length $k$. Thankfully, such entropy satisfies a recursion relation in terms of the Markov transition matrix $M$, $\ ^1 \vec S$ and the parameter $\epsilon<1$,
\begin{equation}
\ ^{k+1} \vec S_\epsilon = \epsilon^{k} \left( \ \frac{M}{\epsilon^{k-1}}\ \ ^{k}\vec S + \ ^1\vec S \right).
\label{eq:rec}
\end{equation}
If we write down all the terms, recursively, we find the more compact formula,
\begin{equation}
\ ^{k+1} \vec S_\epsilon = \sum_{n=0}^{k} \epsilon^n M^n  \ ^1\vec S,
\end{equation}
and, realizing that we can take the limit $k \rightarrow \infty$  safely, we obtain a closed expression :
\begin{equation}
\ ^{*} \vec S_\epsilon = \frac{ 1}{I- \epsilon M} \ ^1\vec S ,
\label{eq:entfin}
\end{equation}
which is finite for all values of $\epsilon<1$, is a fixed point of eqn. (\ref{eq:rec}), and can be interpreted as a non-local notion of entropy. We note that now each node can have different values of the entropy depending on the structure of the Markov transition matrix $M$. The strength of eqn. (\ref{eq:entfin}) is that it is now easier to evaluate the entropy, rather than eqn. (\ref{eq:entropy}). This advantage comes however at the price of a free parameter $\epsilon$, that can be given an \textit{a posteriori} interpretation if we consider this as a probability over the space trajectories. Using this interpretation,  the average length of a trajectory becomes
\begin{equation}
\langle k \rangle = \sum_{k=1}^{\infty} \epsilon^k k=\frac{\epsilon}{(1-\epsilon)^2},
\label{eq:inveps}
\end{equation} 
which can be inverted for any $\langle k \rangle>0$, for $\epsilon\left(\langle k\rangle\right)$.
In this paper we consider $\epsilon$ such that $\langle k \rangle=N$, the total number of nodes of the networks. In general, the limit $\epsilon\rightarrow 0$ recovers the node entropy considered in other works, as for instance in \cite{shortpdv}, of which ours can be considered as a non-local extension. We will study this extended notion of graph entropy to characterize the dynamics of memristive circuits.

It is worth discussing explicitly few properties of the entropy of eqn. (\ref{eq:entfin}). We observe first that the limit $\lim_{\epsilon\rightarrow 1} \ ^{*} \vec S_\epsilon$ is unbounded. This is due to the fact that the Perron root of a Markov chain is $1$ and that thus the resolvent of the operator does not exist. However, eqn. (\ref{eq:entropy}) for finite $k$ is indeed bounded from above by $\epsilon^{k-1} k \log(N)$, which is the extreme value of the entropy for finite $k$, corresponding to a complete graph with $N$ nodes. We note that this entropy goes to infinity in the limit $k\rightarrow \infty$ if and only if $\epsilon\geq 1$. If we set $\langle k \rangle=N$, we can obtain an approximate upper bound for the fixed point entropy of every node by inverting for $\epsilon(N)$ in eqn. (\ref{eq:inveps}), and given by
\begin{equation}
\ ^* S\lessapprox 2^{1-N} N \left(\frac{2 N+\sqrt{4 N+1}+1}{N}\right)^{N-1} \log(N),
\end{equation}
which is, for $N\gg1$, of the order of $S_{max}\approx N\ \log(N)$, which unsurprisingly depends in a non-linear way on the total number of nodes. This means that the maximum of the entropy is not well behaved in the thermodynamic limit ($N\rightarrow \infty$), as $S_{max}/N$ does not exist. As a general feature of entropic measures, observing a decrease in graph entropy implies that some structure is being generated dynamically, rather than converging towards a random configuration. As a last comment, we note that this definition does not lead to a unique definition of transition probability when the network is directed. In this case, one can define the \textit{forward} probability, as the one used above, or the \textit{backward} probability. For a directed adjacency matrix, one has the in-degree and the out-degree, defined as diagonal matrices,
\begin{eqnarray}
D^{in}_{ii}&=&\sum_{j} A_{ij},\\
D^{out}_{ii}&=&\sum_{j} A_{ji},
\end{eqnarray}
and thus one has two definitions of transition probabilities:
\begin{eqnarray}
M^f&=& (D^{out})^{-1} A,\\
M^b&=& (D^{in})^{-1} A^T.
\end{eqnarray}
Both these definitions will be used in the following.

The entropy introduced in this section contains topological informations about the graph flows. Thus, it has the advantage of being \textit{non-local} and with support on the nodes of the graph.

\section{Results}

\subsection{Toy diffusion model}\label{sec:rdcomp}

We analyze the topological properties of evolving graphs under the reinforcement-decay model using the graph entropy of Section \ref{sec:gent}.

\begin{figure}[b!]
\begin{center}
\includegraphics[width=8cm]{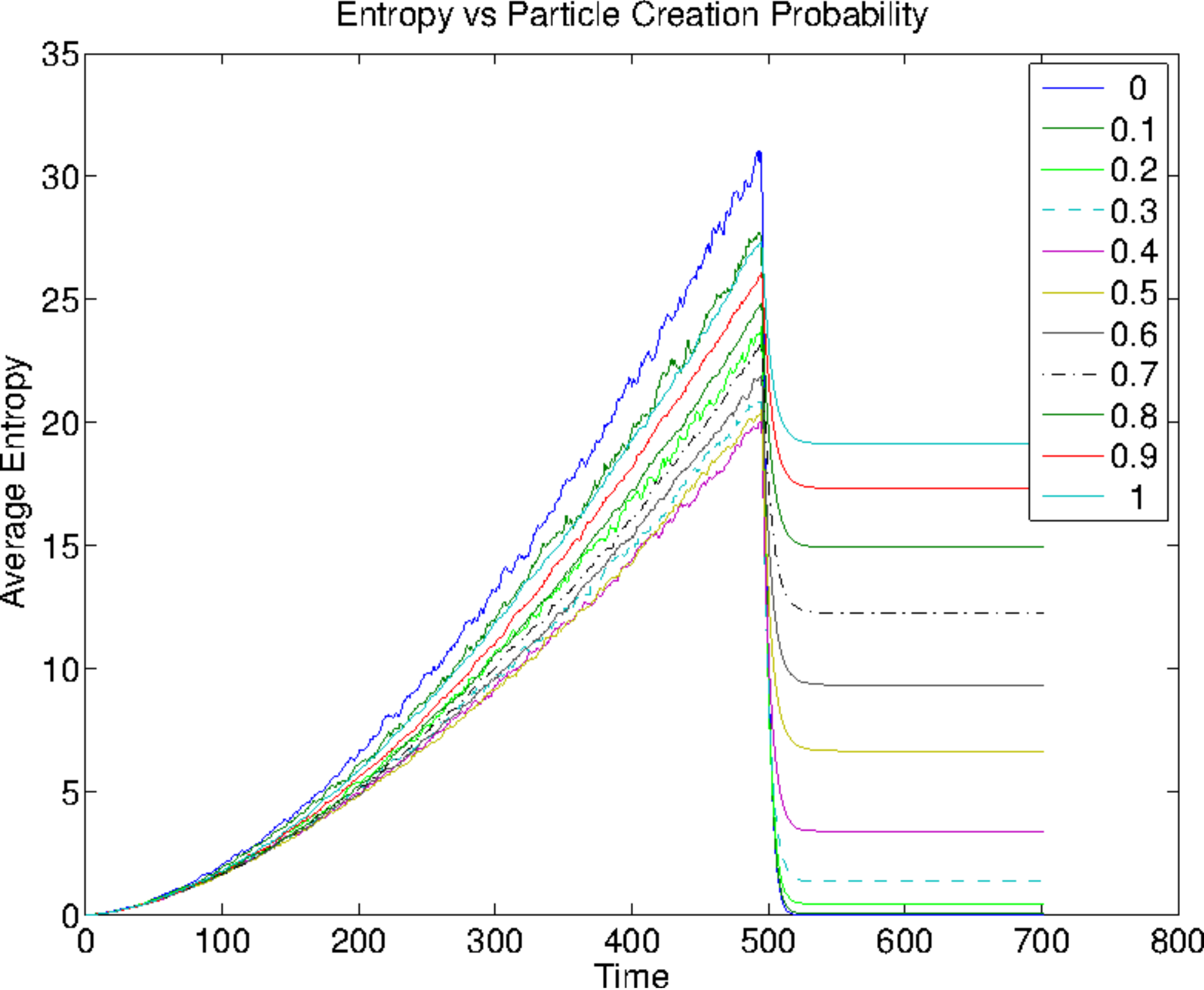} \includegraphics[width=8cm]{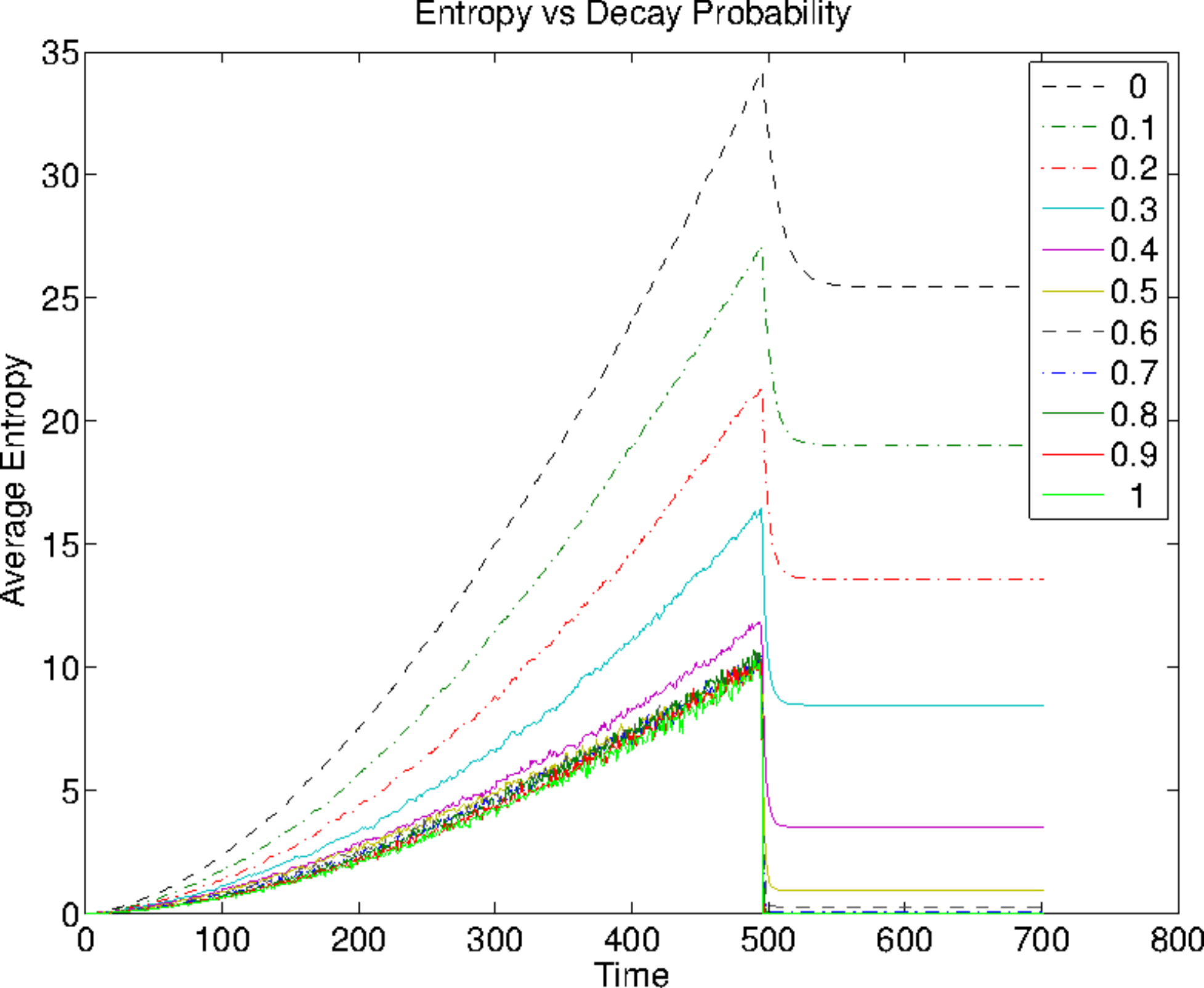}
\end{center}
\caption{Evaluation of the mean entropy as a function of time in the reinforcement-decay model. The growth of the graph stops at $T=500$, and afterwards the network converges to a fixed point.}
\label{fig:EPCP2} 
\end{figure}

In particular, we simulate the system for various values of the two main parameters: the particles creation probability $p_p$ and the decay probability $p_d$. We plot the values of the entropy as a function of time in Fig. \ref{fig:EPCP2}. These were obtained after averaging over 30 simulations, for fixed parameters $\gamma=0.1$, $L_d=1$. At each time step, the strength of links are decreased with probability $p_d$ of a value $\gamma$ which we set to $\gamma=0.1$. The graph grows at each time step up to a maximum number of nodes $N_c$, here chosen to be 500, and then the links which did not reach the critical stable value ``evaporate". The new link parameter $p_{nl}$ is set to $0.1$ in the simulations. For each new node, a new particle is added to the graph at the new node with probability $p_p$.  The growth halts artificially to observe how the graph evolves afterwards, and to see the behavior of the entropy. In Fig. \ref{fig:EPCP2} we show the evolution of the graph entropy as a function of time for different the particle creation probability (left) and the decay probability (right).

As a first comment, we note that as the graph grows the mean graph entropy, defined as $\langle S \rangle = \frac{1}{N} \sum_{i=1}^N (\ ^*S_\epsilon)_i$ (i.e. the average over the nodes), grows as well. Although the mean of the entropy does not have a thermodynamical interpretation, this gives a succinct graphical representation of the outcome.
The growth is in general due to both the decrease in the number of nodes and to the lack of structure due to the random growth. In addition, we stress the important effect of the decay, which competes with the reinforcing of hopping particles. Meanwhile decay decreases the amount of order, particles create structure. We note that after the growth stops, the entropy decreases to a stable value. This implies that the effect of decay is of creating ``structure" by removing most of the unstable links. 

Notably, we observe that for each set of parameters, the graph entropy curve takes different values, both during the growth and after. This implies that such graph entropy measure is enough to characterize the asymptotic state of the graph in the simulations we have performed. Thus, in all the performed simulations, the average entropy per node is indeed characterizing the parameter space.

%\subsection{Purely Memristive networks}\label{sec:memrcomp}

\begin{figure}
\begin{center}
\includegraphics[height=7cm]{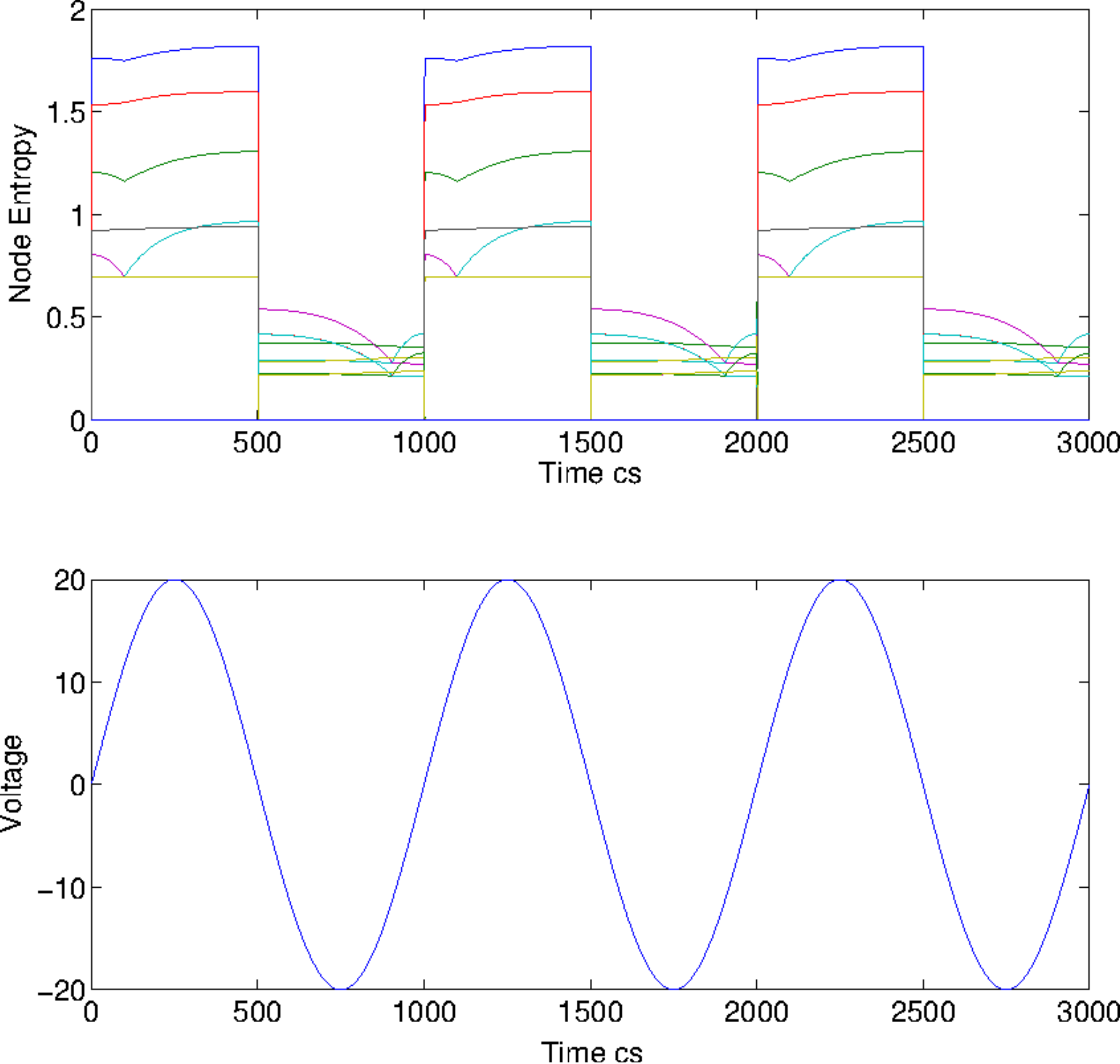} \includegraphics[height=7cm]{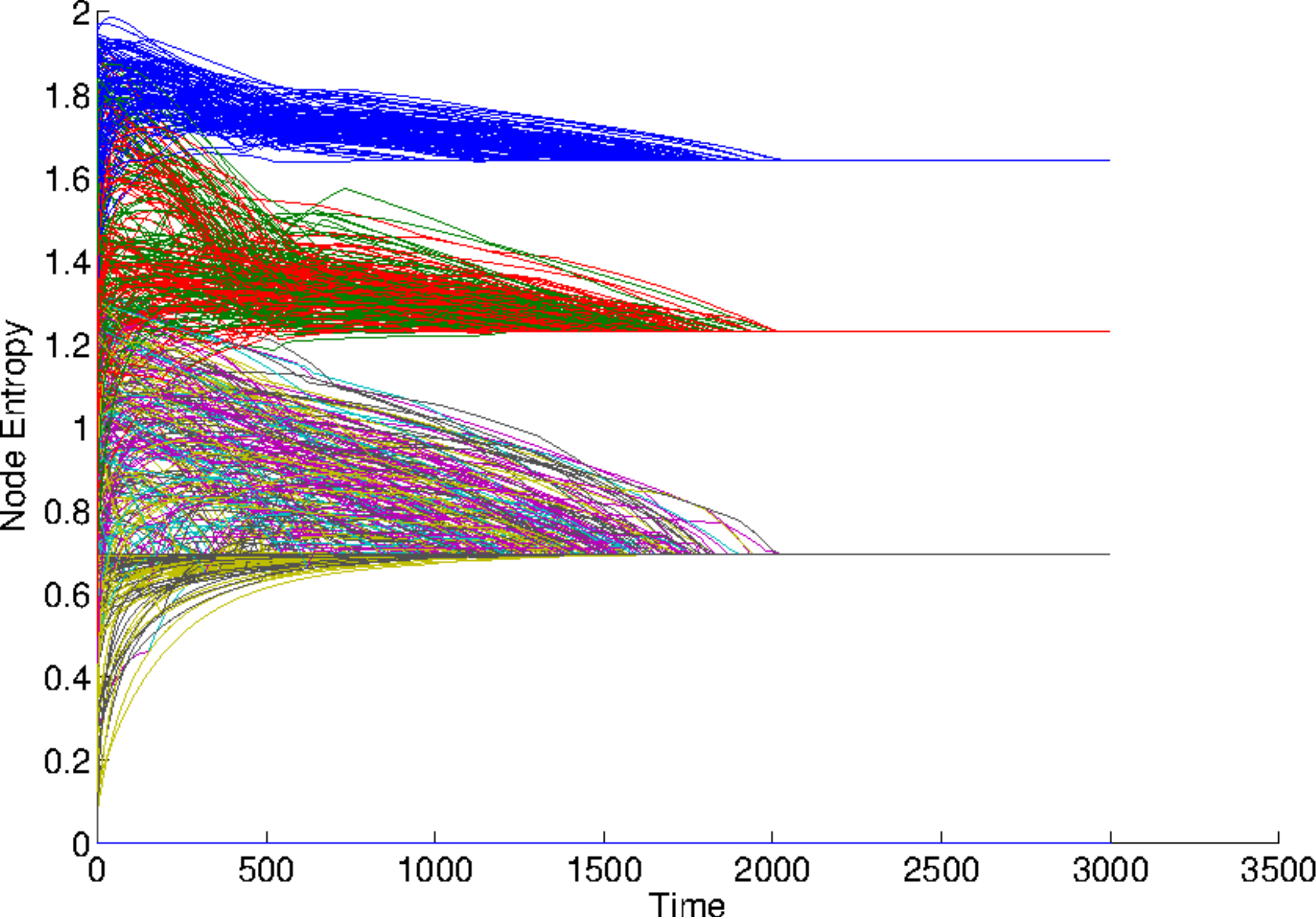}
\end{center}
\caption{ Evolution of the node entropy for AC and DC voltage, measured in centiseconds for random initial conditions. \textit{Top:} Evolution of the entropy for the tree graph of Fig. \ref{fig:TREE}  when an AC voltage is applied to the network with $V=10 \sin(10 t)$ volts. \textit{Bottom:} Entropy evolution for the tree graph of Fig. \ref{fig:TREE}  for a DC voltage $V=60$ volts. The Monte Carlo simulation with all the 100 generated trajectories for $L=4$, with $V=60$ volts. We note that all the trajectories converge to the same asymptotic fixed point.}
\label{fig:TREEAC}
\end{figure}

\subsection{Purely Memristive circuits}\label{sec:memrcomp}

In the previous section we observed that the non-local graph  entropy provides characterizes the dynamics of the reinforcement-decay model.  We thus use the entropy to study the dynamics of purely memristive circuits. 

In order to apply the entropy to the case of memristive circuits, we consider the following mapping between a flow on a network and a stochastic matrix. Given a network with a unique directed edge between two nodes, we introduce the matrix $I_{ij}$, given by the flow of currents between nodes $(i,j)$, and defined as a matrix which is elementwise non-negative. For each couple of nodes $(i,j)$ connected by a resistance or memristance, only one of the elements of $I_{ij}$ or $I_{ji}$ are non-zero, depending on the directionality of the current. Since $I$ is a non-negative matrix, we can obtain a stochastic matrix by normalizing by rows (columns), obtaining $\tilde I$, satisfying $\sum_{j} {\tilde I}_{ij}=1$ ($\sum_{i} {\tilde I}_{ij}=1$). A notion of entropy on the set of currents was introduced in \cite{shortpdv} in order to study the self-organizing properties of memristive circuits. Here we intend to extend this analysis to understand non-local spatial properties using the graph entropy measure of the previous section. 

In order to simulate the dynamical evolution of the memristive circuits, in the present paper we use nodal analysis \cite{aoe} to solve for the currents at each time step, and a Runge-Kutta 4 method to implement the dynamics. We then construct the Markov matrix from the resulting currents, and calculate the node entropy dynamically.

A particular feature we aim to explore is the role of symmetry. As previously described, Fig. \ref{fig:TREE} we show the triangular network with $L=3$ layers, which can be easily generalized to a higher number of layers. In each layer, there are both horizontal memristors closing on a triangular mesh, and vertical memristors connecting to a new layer. For a triangular network with $L$ layers, there are $3*(2^{L}-1)$ memristors to simulate. In the last layer, the horizontal memristors are not placed, as these have zero voltage difference applied to them due to the connection to the ground.
 Let us consider first the case in which there is no state decay, $\alpha=0$, and in which all the polarities of the memristors are aligned alike $p=1$. In particular, we simulate  homogeneous memristors parameters, with $R_{on}=100$ Ohm, $R_{off}=16000$ Ohm, $\mu=10^{-3}$, $d=10^{-11}$, and with a time step given by $dt=10^{-2}$ s$=cs$ and with constant external voltage $V=20$.

As a first example, we consider the case $L=4$ under the forcing of an AC voltage generator, with  $V(t)=20 \sin(10 t)$. In Fig. \ref{fig:TREEAC} we show the entropy evolution for each node (left panel - top) and the applied voltage (left panel - bottom), for a random initialization of the memristors state. It is easy to see that the negative and positive voltage sides of the generator correspond to different phases of the circuit. In fact, the node entropy changes dramatically and periodically with time, following the change of currents signs. The case of a DC generator is shown in Fig. \ref{fig:TREEAC} (right panel) for an initial random configuration. Interestingly, from the entropy we can identify the 4 different entropy layers. We note that initially being each memristor randomly initialized, each node has different entropy level. Dynamically however, the non-local entropy measure is sensitive to changes to far nodes of the network, as it can be easily observed. Each entropy level reaches an asymptotic value which depends on the layer only. The fourth layer has  zero value for each node. This is given to the fact that each node is connected to the ground, and only a unique current is connected to each node.

\begin{figure}
\begin{center}
\includegraphics[width=5.5cm]{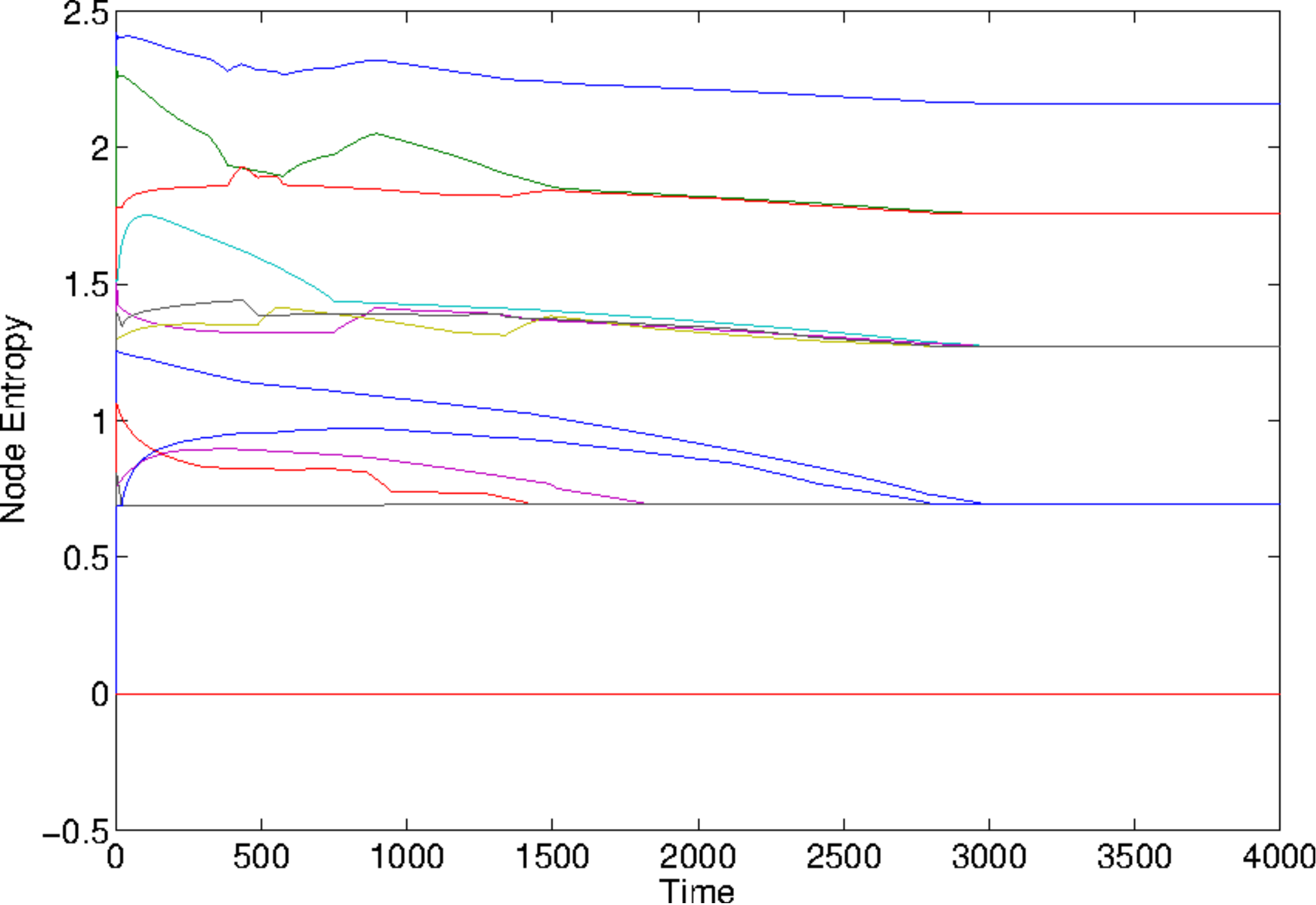} \includegraphics[width=5.5cm]{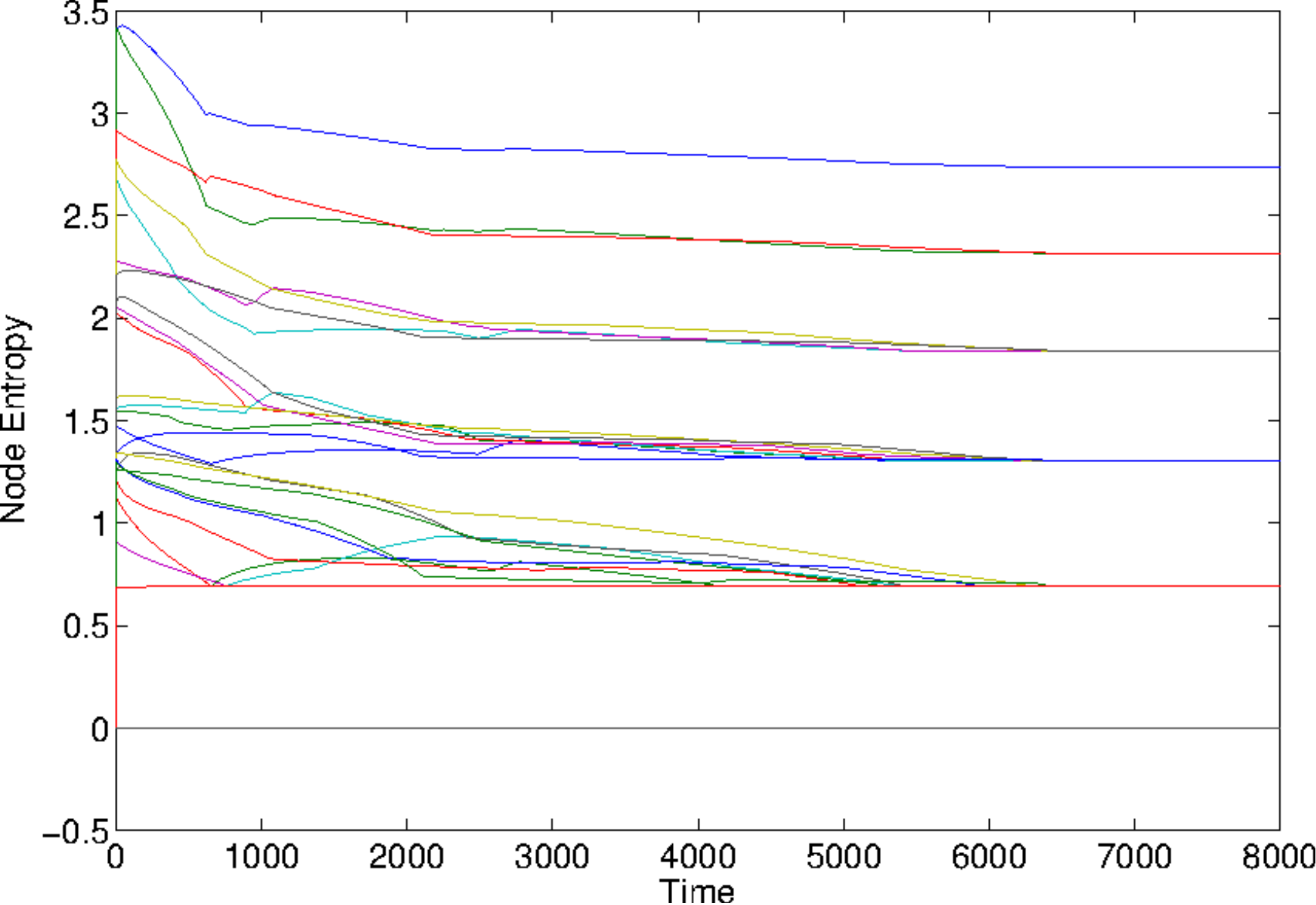}\includegraphics[width=5.5cm]{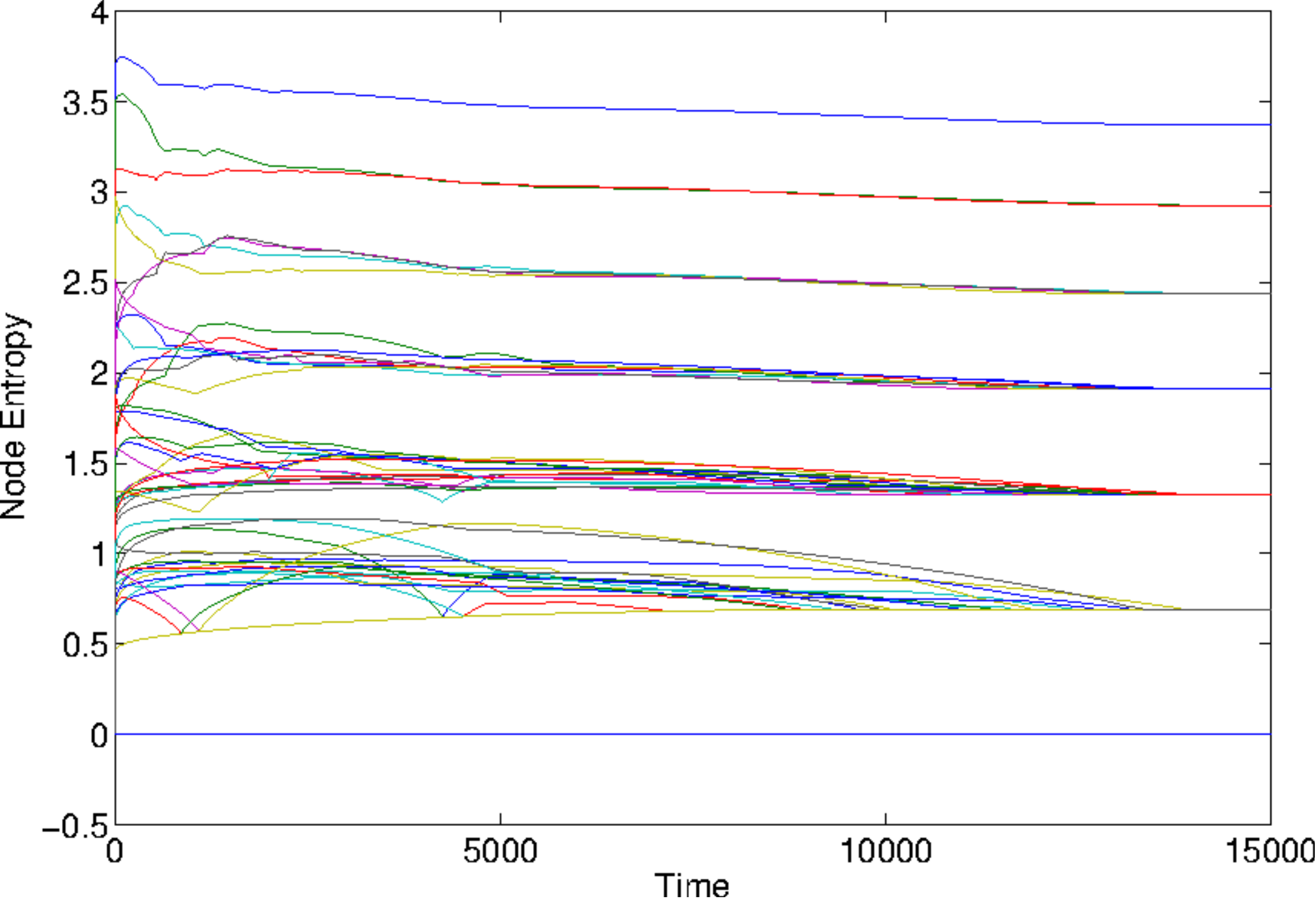}
\end{center}
\caption{Graph Entropy for the tree circuit of Fig. \ref{fig:TREE} for L=5 (left), L=6 (center) and L=7 (right), in the case of DC generator, with $V_0=60$ volts. The initial configurations are chosen at random. We observe the emergence of $L$ layers of entropy, showing that the system self-organized in order for the nodes in each layers to have equal entropy.}
\label{fig:ETREE}
\end{figure}

A similar  behavior is observed also for the case of a higher number of layers, as shown in Fig. \ref{fig:ETREE} for the cases $L=5, 6$ and $7$.  Few peculiar things can be noted. The first observation is that the number of layers is equal to the number of asymptotic levels of the entropy, and the number of nodes with that level of entropy corresponds exactly with the number of nodes in that layer (for the $L$-th layer, there are $2^{L-1}$ nodes). Moreover, we note that the time it takes for the entropy to reach its asymptotic value increases with the number of layers. Visually, one can observe that it is $\approx 20$s for $L=4$, $30$s for $L=5$, $\approx60$s for $L=6$ and $\approx120$s for $L=7$. This is confirmed in Fig. \ref{fig:EHIST} by calculating the histogram of the asymptotic entropy value for the cases $L=4,5,6,7$.

\begin{figure}
\begin{center}
\includegraphics[width=8cm]{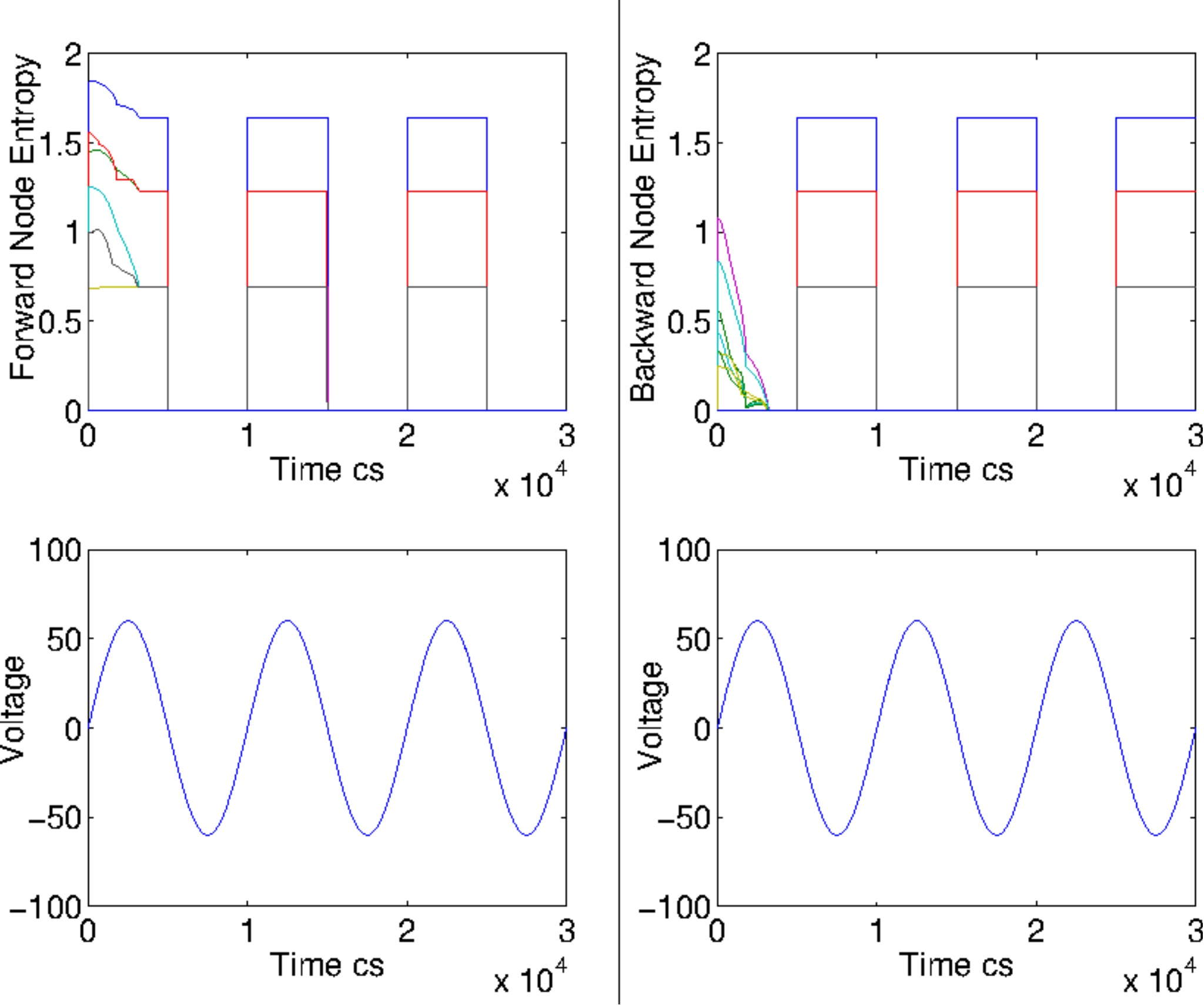}
\end{center}
\caption{We show the difference between the node complexity measure based on forward probability and backward probability in the case of AC controlled memristive circuits. If the frequency is lower than the inverse of the half thermalization time, then the entropy is zero in the negative voltage part of the generator, as the left plot shows. In this case, one has to use backward probabilities, as in the right plot.}
\label{fig:FB}
\end{figure}

We observed numerically that, independently of the initial condition, the thermalization time increased with the number of layers; we define the thermalization time as the time it takes to the entropy to reach its asymptotic and stable value. Specifically, heuristically we observed that if the frequency of the generator in the AC mode is longer than the inverse of twice the thermalization time, then in the negative voltage part of the generator, the node entropy becomes zero. This is shown for instance in Fig. \ref{fig:ETREE} for the case of $L=4, L=5$ and $L=6$, and where the frequency has been tuned to be larger than the observed thermalization time.
In the negative side of the generator, the system can be described by, instead of the forward probabilities, by the backward probabilities, implying that the stochastic matrix is derived from the transpose of the current matrix $I_{ij}$. The difference is shown in the plot of Fig. \ref{fig:FB}, which illustrates that the positive voltage of the generator is characterized by the forward probabilities matrix, but the negative side is characterized by the backward probabilities.

Our analysis has important implications. In particular, we observe that asymptotically the memristive circuit considered here converges to a state which respects the symmetry of the network. The fact it converges to a specific value, although surprisingly unique and robust, can be thought as a feature of the dynamical system. 
Moreover, since it starts from random configuration of the memristors, this implies that the symmetry-respecting state is a robust attractor of the system and with a large basin of attraction. We also observe the network rearranges itself in order to make superfluous some of the memristances. It is easy in fact to realize that since the nodes in a layer must be equal, the memristances arranged horizontally and belonging to the same layer are connected to nodes of equal voltage. This implies that no current is flowing in these memristances.
\begin{figure*}
\begin{center}
\includegraphics[width=12cm]{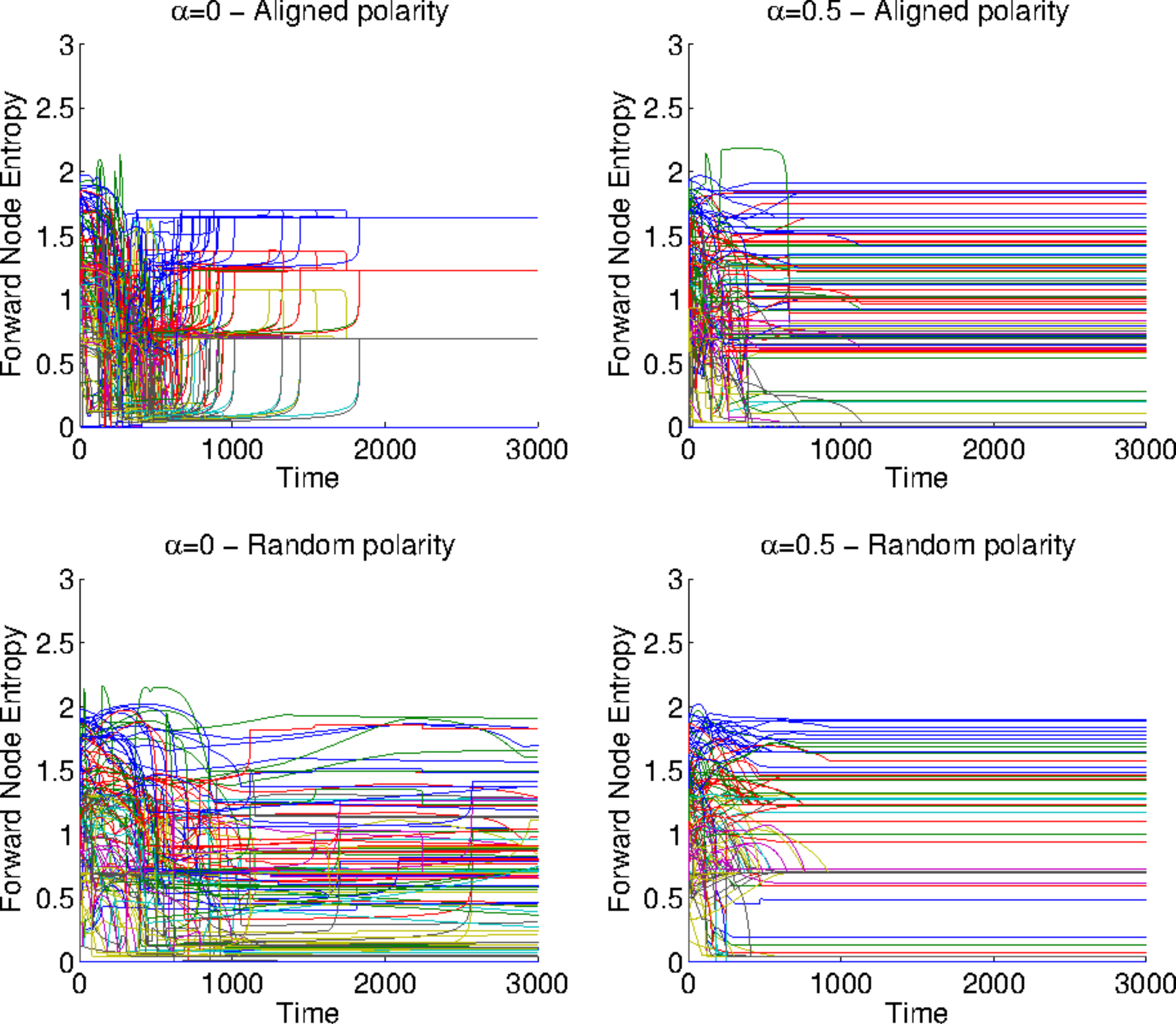}
\end{center}
%\
\textbf{\refstepcounter{figure} \label{fig:ETREEPD} Figure \arabic{figure}.}{  Comparison between simulations without state decay and aligned polarities (top left), only state decay $\alpha=0.5$ (top right), randomized polarities (bottom left), and randomized polarities with state decay (bottom right). The input voltage is $V_0=50 V$. The plots were obtained from plotting all the results from 100  Monte Carlo simulations with similar dynamical parameters but where the initial states have been randomized.}
\end{figure*}

An extension of the model above can be made to account for both state decay and different polarities. In order to check whether the state is truly symmetry respecting, we simulate for the case $L=4$ with the polarity assigned at random and with equal probability. In the top row of Fig. \ref{fig:ETREEPD} we show the case with aligned polarities for $V=50$ Volts. 

In Fig. \ref{fig:ETREEPD} (top left) we show the case without state decay for constant $V=50$ Volts, $\alpha=0$, for comparison with the case $\alpha=0.5$, which is shown in Fig. \ref{fig:ETREEPD} (top right). In general, we observe that both state decay and switched polarities removes the symmetry respecting properties of the dynamics.

The bottom row of Fig. \ref{fig:ETREEPD}  shows the case with randomly assigned polarities, both for the cases without (bottom left, $\alpha=0$) and with (bottom right, $\alpha=0.5$) state decay. The difference with the top left picture is notable. Both in the cases with switched polarities and decay, the network symmetry is lost, as now each memristor can have two different polarities. 

\begin{figure*}
\begin{center}
\includegraphics[width=16cm]{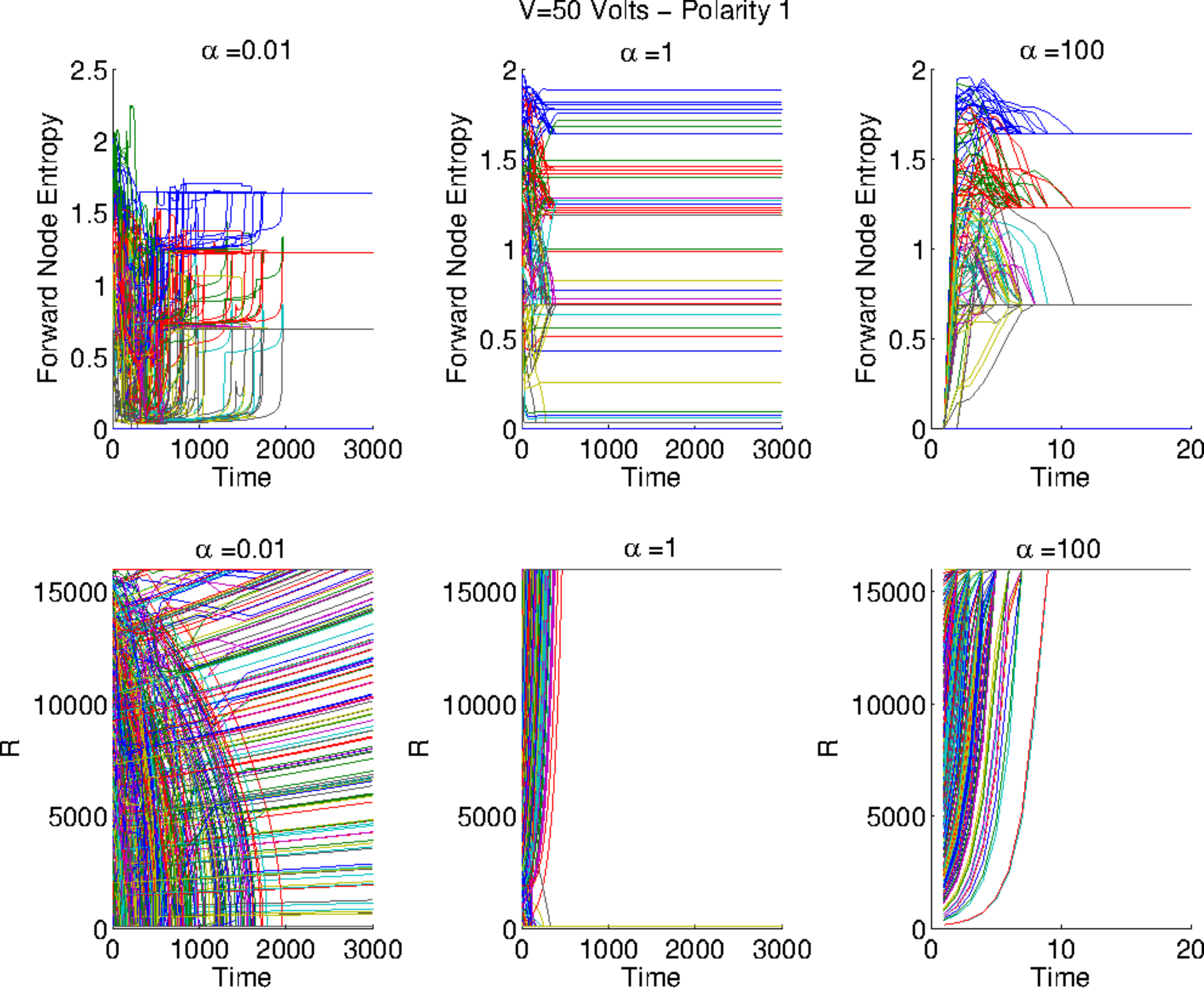}
\end{center}
%\
\textbf{\refstepcounter{figure} \label{fig:RESIST} Figure \arabic{figure}.}{  \textit{Left:} Evolution of the resistances and the entropy for various values of the decay parameter $\alpha$ and for input $V=50$ Volts.  \textit{Left column:}  Case with $\alpha=0$. We see that the resistances take different final values , but that the entropy converges to some specific values. \textit{Central column:} Case with $\alpha=0.5$. For this intermediate value of $\alpha$, we observe that some memristances decay to the $R_{off}$ value, meanwhile some resistances flow to the $R_{on}$ state. In this case, we observe that the entropy can take different values.  \textit{Right column:} For $\alpha=100$, all resistances quickly onverge to the $R_{off}$ value. However, we observe an analogous entropy configuration to the one of the case $\alpha=0$.}\end{figure*}

However a careful analysis shows that this not true for all values of the decay constant, as we see in Fig. \ref{fig:RESIST}. We observe that the state decay does not modify the asymptotic value of the entropy for small values ($\alpha\approx 0.01$) and large values ($\alpha\approx100$) of the decay constant; although the behavior of the resistances is different in the two cases, as we show in Fig. \ref{fig:RESIST} (left and right). We observe however the dynamics of the entropy is clearly different, showing that the two models are inequivalent. The case of intermediate ($\alpha\approx 1$) values of the decay constant are shown in Fig. \ref{fig:RESIST} (center), a dependence on the initial condition is present. In general, in this case the resistances converge either to $R_{on}$ and $R_{off}$ asymptotically, and the entropy is observed to take different values from those observed for $\alpha=0$.

\begin{figure}
\begin{center}
\includegraphics[width=8cm]{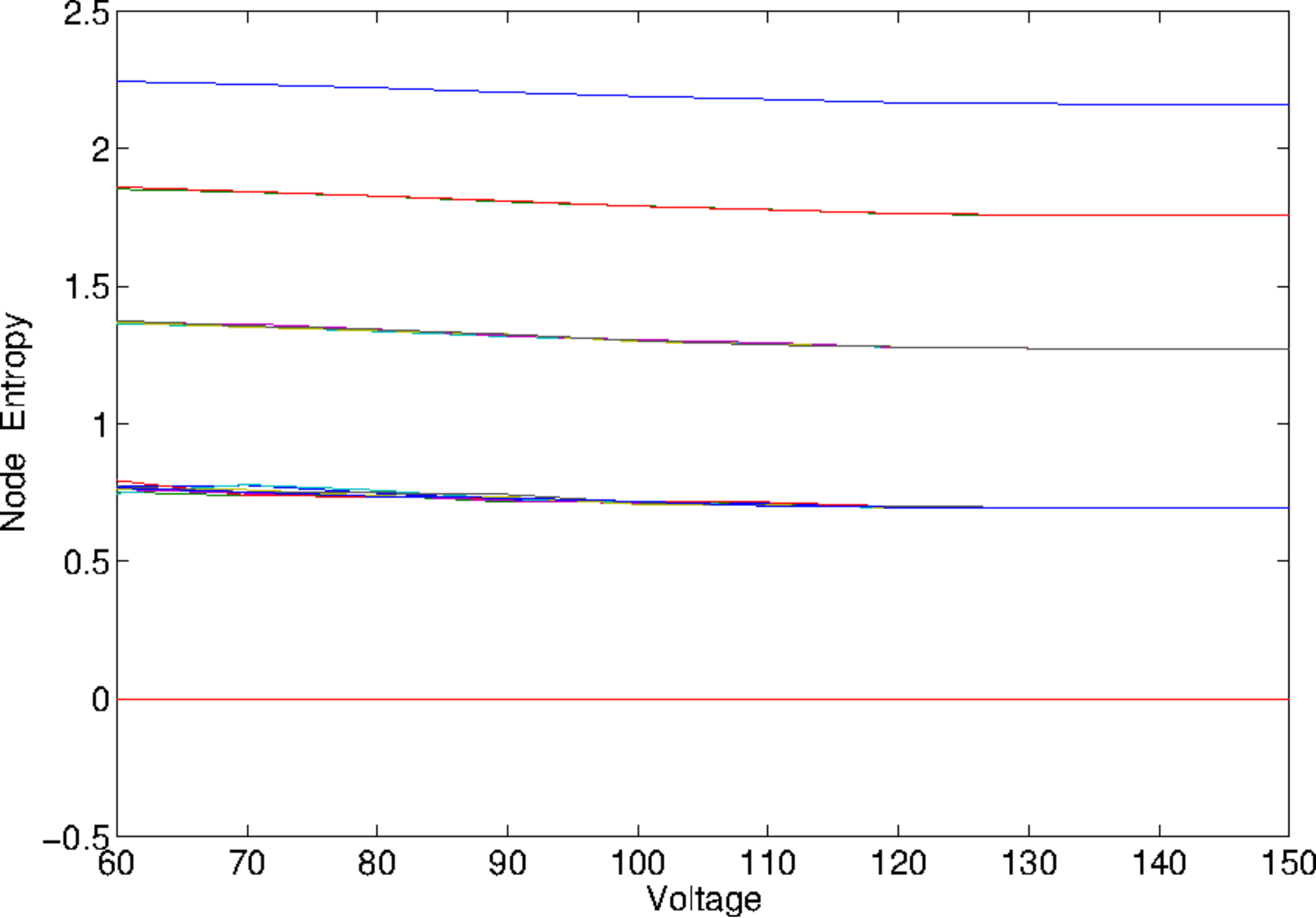} \includegraphics[width=8cm]{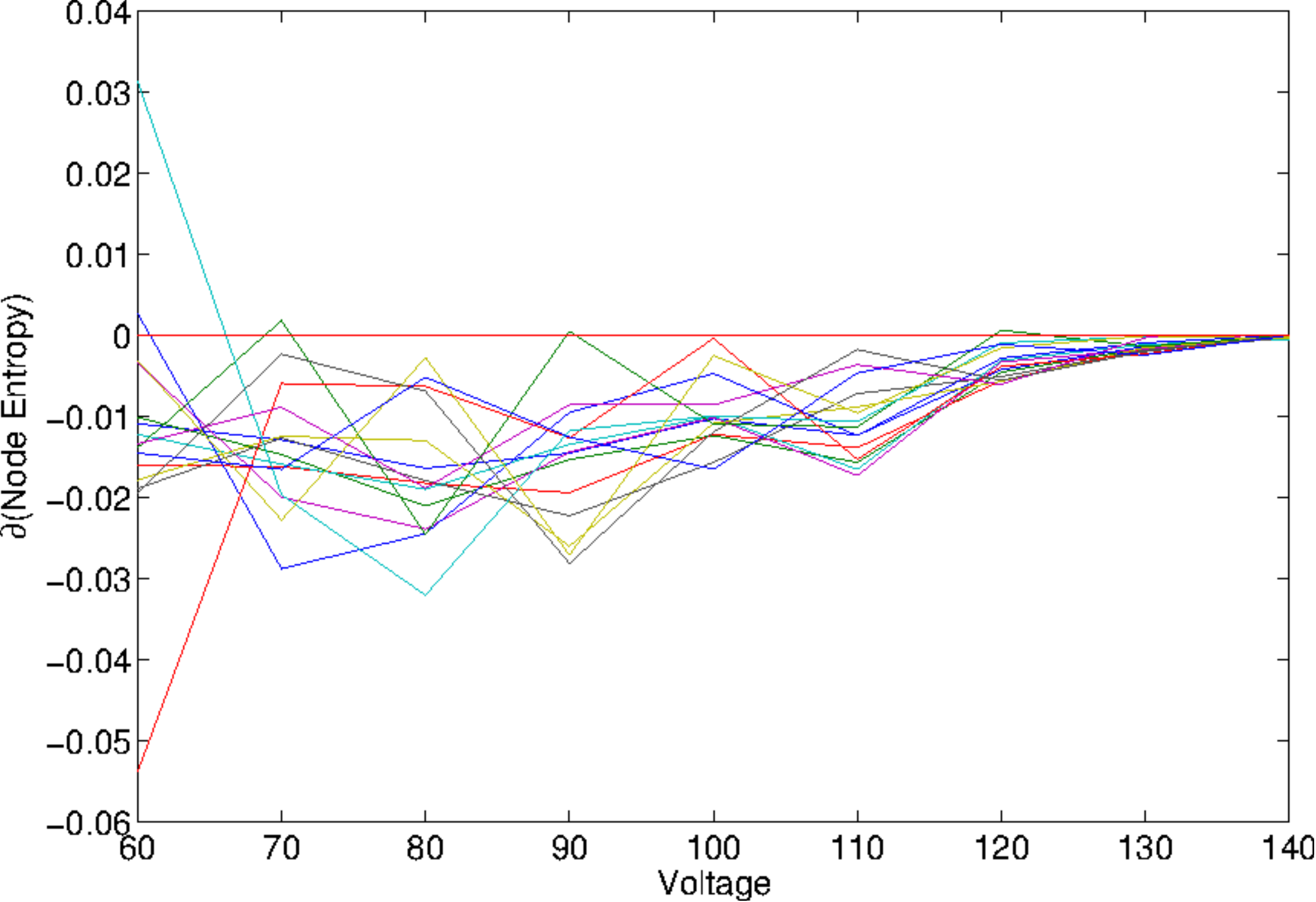}
\end{center}
%\
\caption{  Average Node entropy (left) evaluated from a Monte Carlo simulations over $100$ instances, in the case $L=5$. The voltage is varied from $V=60$ to $V=150$ in steps of 10 volts. On the left, we plot the derivative of the entropy wrt the input voltage. We observe that for large values of the voltage, the derivative converges to zero. Multiple simulations have shown that the value of $V$ where the derivative crosses zero is not universal, but indeed varies with the parameters of the memristors.}
\label{fig:ETREE1}
\end{figure}

 To conclude, we provide a sensitivity analysis of the graph entropy as a function of the applied voltage in the case of a DC generator, no state decay and aligned polarities. In Fig. \ref{fig:ETREE1} we plot the asymptotic entropy values for each node in the case $L=5$ (left), and the derivative with respect to the voltage (right). The plots have been obtained after a Monte Carlo, and averaged over 100 initial random initialization of the memristor internal parameters. We note that there is a value of the voltage applied by the generator, which we identify as $V_c$, after which the asymptotic entropy does not change. Although we stress that the actual value of $V_c$ depends on the parameters characterizing the memristors, after varying the parameters we find a similar behavior. In Fig. \ref{fig:ETREE1} (right) we see that the derivative of the entropy becomes zero at $V_c\approx 150\ V$. Further analysis, beyond the scope of this paper, is required to understand whether this transition presents any form of critical behavior.

\begin{figure*}
\begin{center}
\includegraphics[width=6cm]{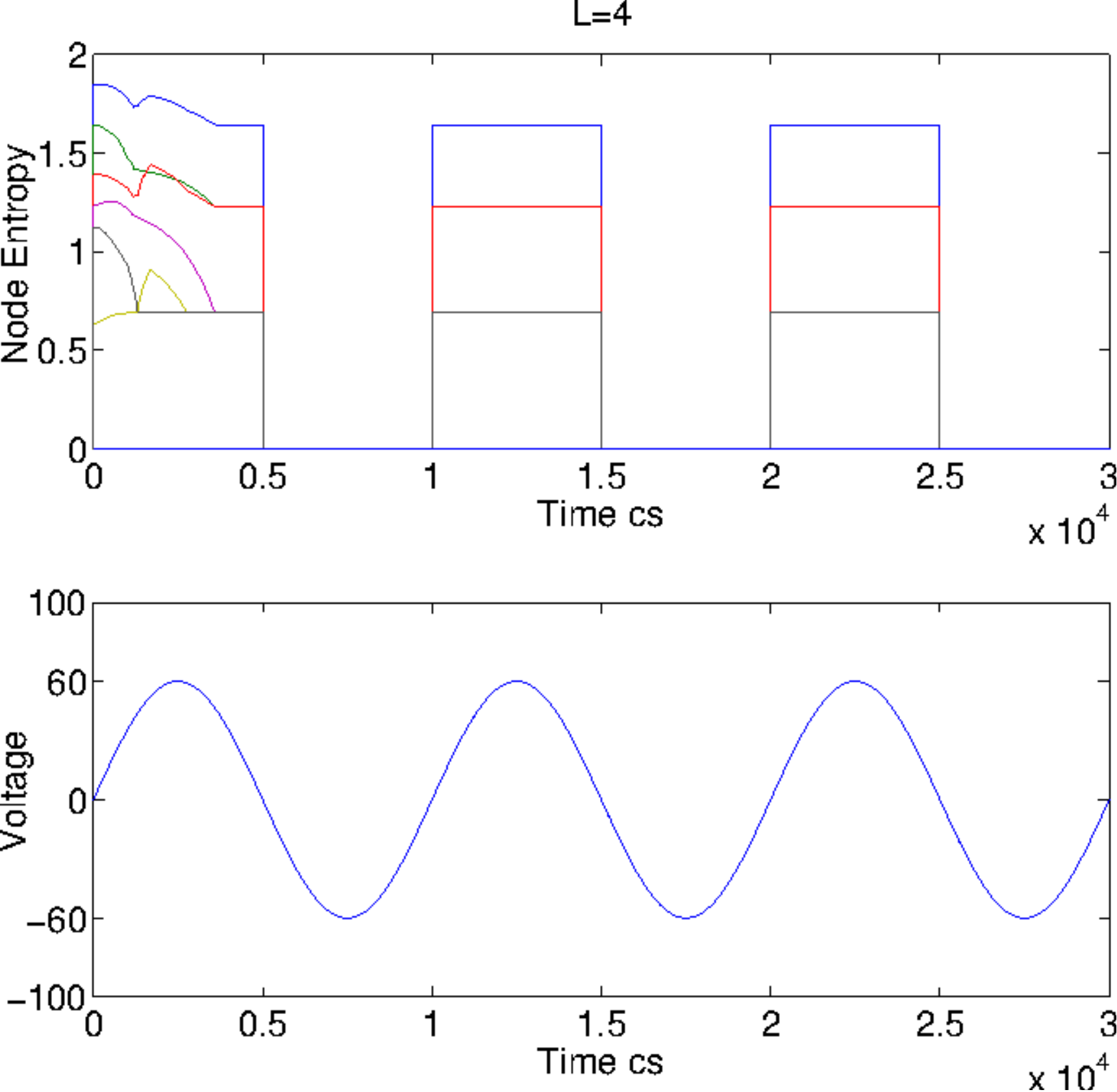}\includegraphics[width=6cm]{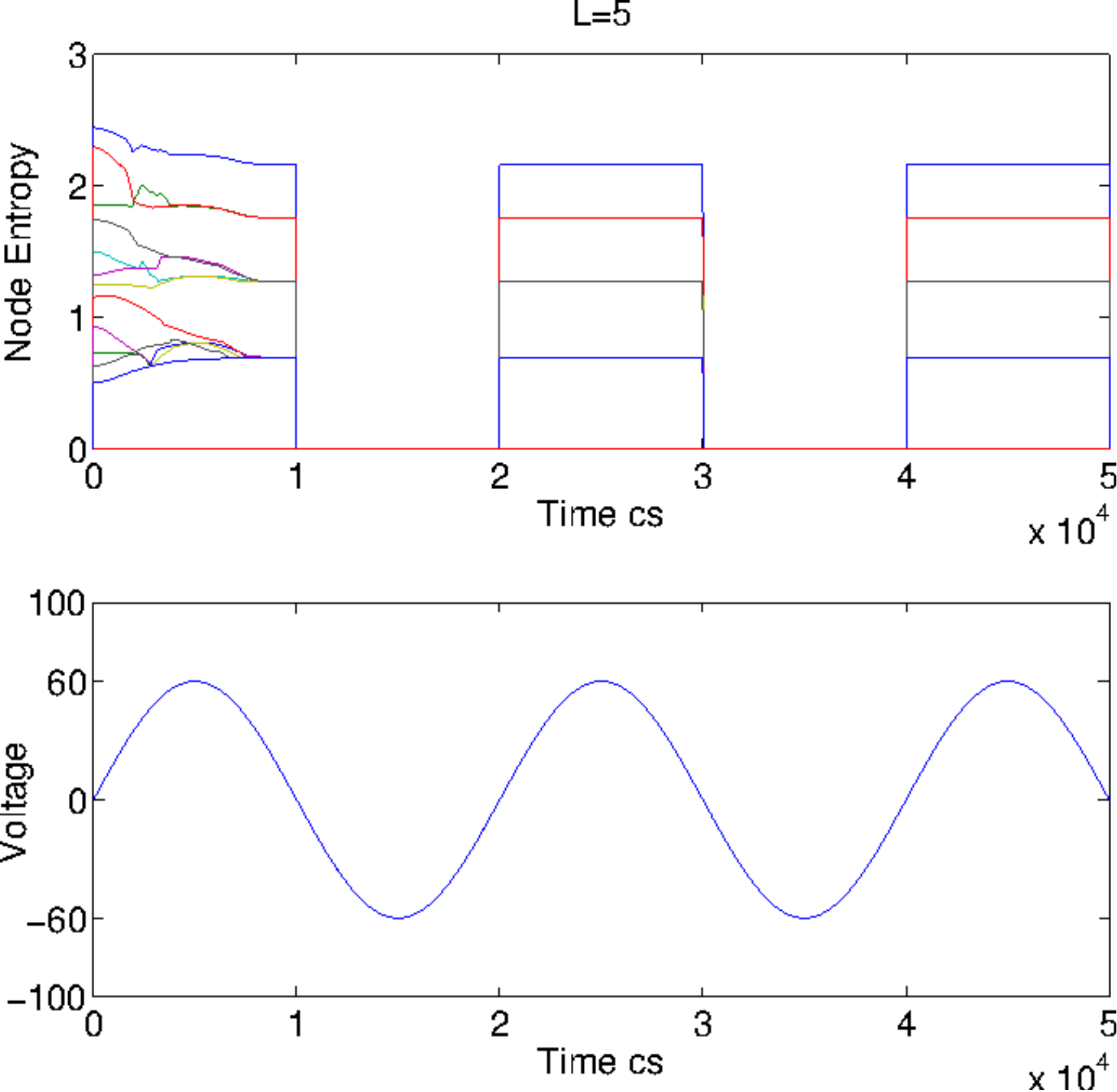}\includegraphics[width=6cm]{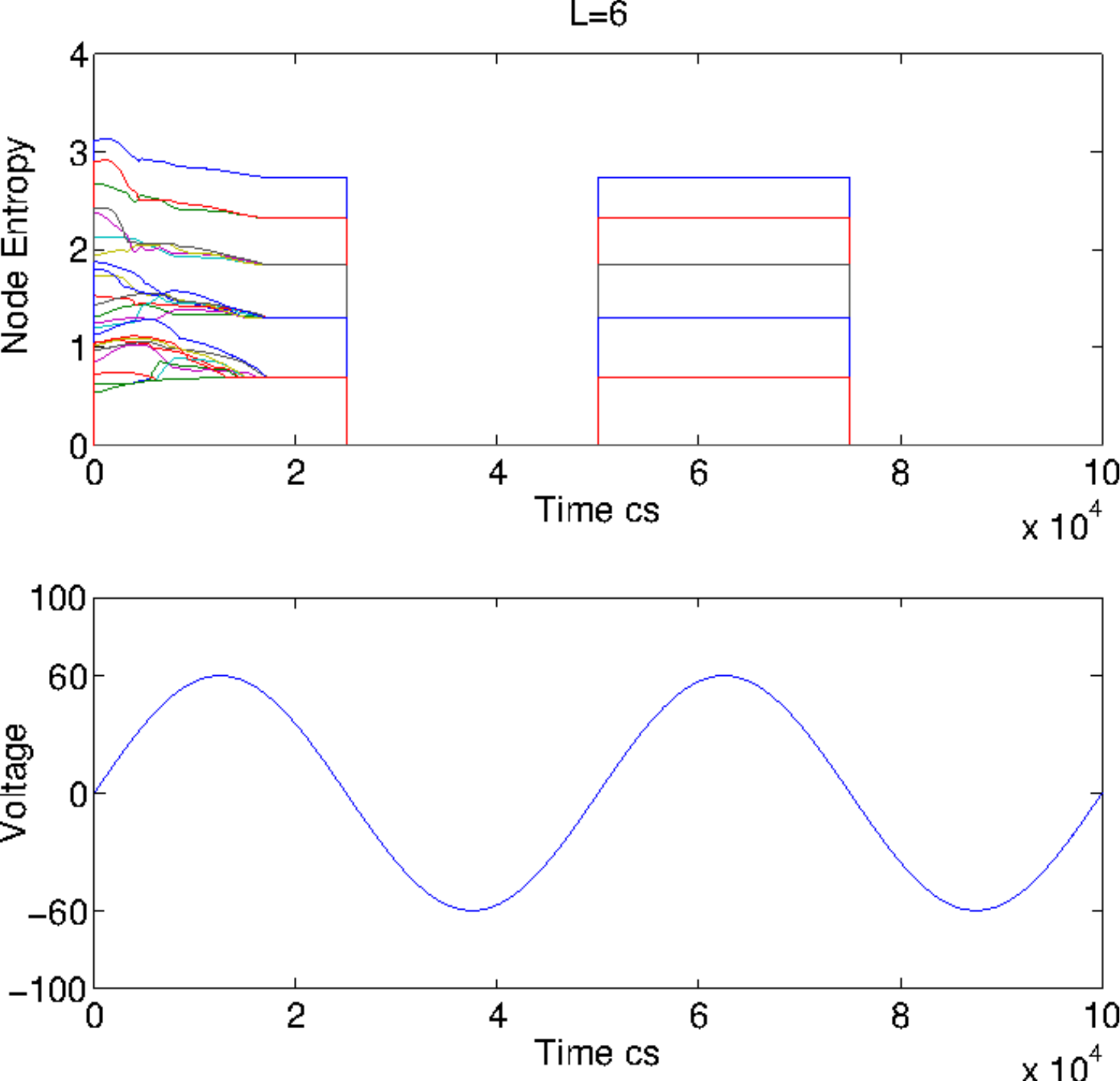}
\end{center}
\caption{Node complexity based on forward probability for the case of $L=4$, $L=5$ and $L=6$ networks, for a single instance from random initial configuration of the memristors. The frequency has been tuned to let the system thermalize in the upside of the voltage generator. We see that in the negative part of the voltage generator, the entropy becomes zero.}
\label{fig:ETREEEX} 
\end{figure*}

\section{Discussion}
In this paper we applied the entropy measure introduced in \cite{rankingcar} to study the dynamics of evolving graphs with memory. In particular, we studied the toy model introduced in \cite{our}, inspired by evaporating ant pheromone trails, a process known to be able to solve problems as finding the shortest path between their nest and food. The pheromone track has a characteristic decay time, but is reinforced every time by ants. We have also applied the non-local entropy measure to the case of pure memristive circuits \cite{williams,chua}.  A local version of the entropy used in this paper was in fact introduced in \cite{shortpdv} to study the self-organization properties of memristors. This non-local extension can be thought of as the centrality operator applied to the local definition of entropy of a node in a graph and depends on an external constant which controls the amount of non-locality. We fixed this parameter to allow loops large enough to enclose all the nodes in the network. This also fixes the maximum entropy one can expect from a node, and we have provided a formula for this extreme value. 

The evolution of the node entropy was studied numerically both for the reinforcement-decay model and for the case of memristive circuits in a tree-like structure. We considered these two models and their relaxation to the asymptotic entropy values for each node. Although our results not being of general character, we have observed 
that the non-local entropy measure studied here characterizes the asymptotic graph structure. For the case of the reinforcement-decay case, we have found that for the entropy distinguishes dynamically different evolutions with different parameters.

The same analysis has been performed for the case of memristive circuits, and applied to the specific case of tree-like structures which can be defined recursively, and shown in Fig. \ref{fig:ETREE}. In fact, in general the entropy reveals that the memristive circuit self-organizes, reaching a configuration in which each node in the $k$-th layer has the same entropy. We distinguished  the case of AC and DC controlled circuits. In the case of AC circuits, we have noted that there are two regimes: the one in which the frequency is larger or smaller than the inverse of half the thermalization time. 
 If the frequency in the AC controlled circuit is low enough, the circuit thermalizes, and in the negative voltage side of the sinusoid the entropy of the network becomes zero. We have shown this is an artifact of the entropy, and that it is necessary to consider both forward and backward probabilities in the definition of the Markov transition matrix. 
 
 \begin{figure*}
\begin{center}
\includegraphics[width=14cm]{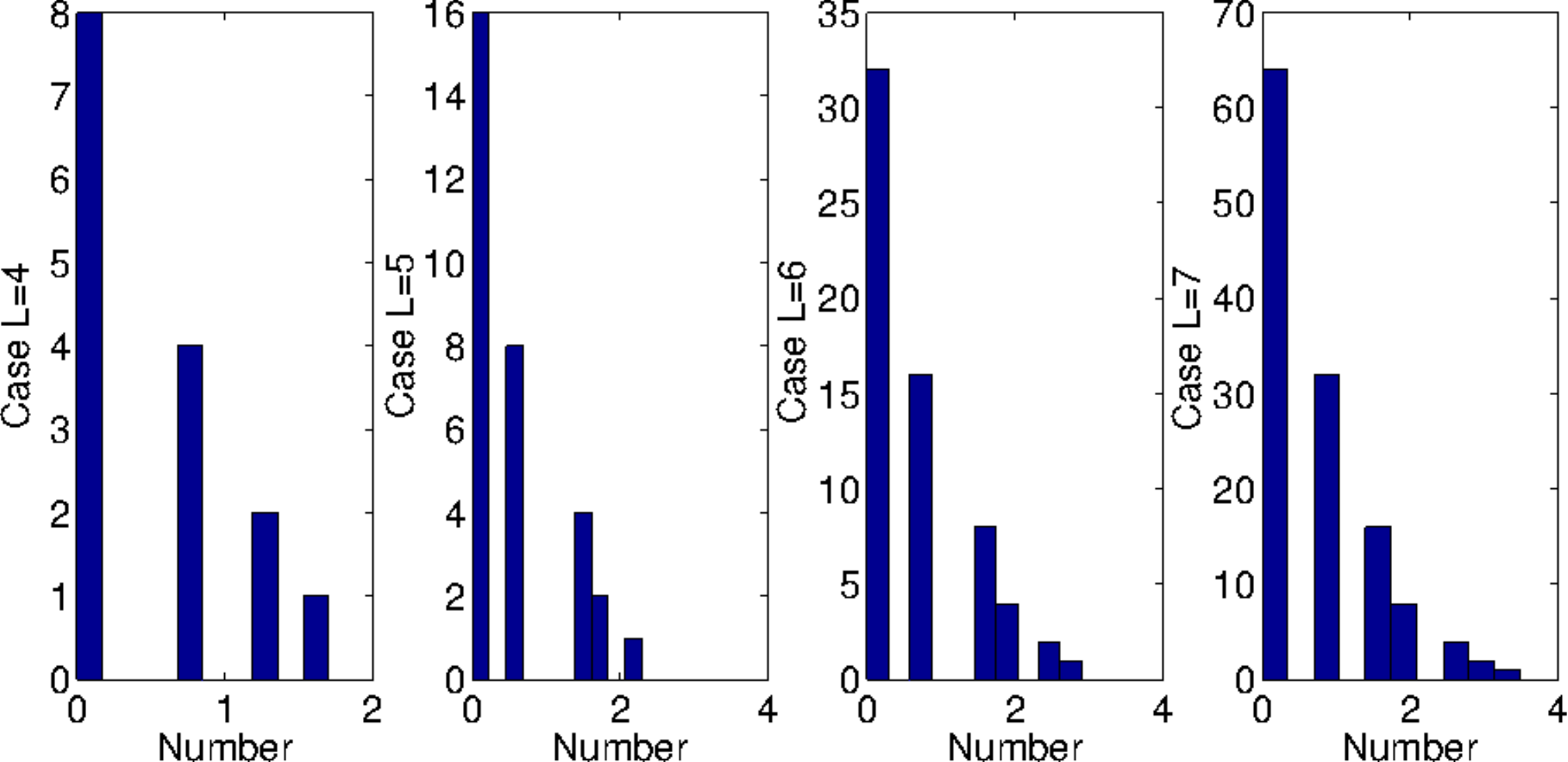}
\end{center}
\caption{
 Histograms of the asymptotic entropy values for $L=4,5,6,7$. We observe that the maximum entropy increases, and that the number of values on each level grows as $2^L$.}
\end{figure*}

An observed feature is that the network re-organizes to respect the graph symmetries. We observed that this statement is true as long as the assigned polarities are aligned and one does not consider the effect of state decay.
We considered the case in which the polarities of the memristors were chosen at random, and the dynamics of the internal parameter has been extended to consider state decay. In both cases the symmetry of the graph is destroyed and thus one observes that the asymptotic entropy configuration can take any value depending on the initial condition. This has been observed to be true only for values of the decay constant of order one, but for small values the self-organizing properties of memristors persists.

We have shown that in general the asymptotic level of the entropy for each node depends, in the DC case, on the voltage applied. We have seen that if the voltage is high enough, the entropy converges to a fixed value. This value changes depending on the parameter of the model, and thus is not universal. The analysis of this phenomenon goes beyond the scope of this paper, and will be analyzed in future works, but shows that the entropy considered in this paper has interesting properties.

The findings of this article  suggest that non-local measures of self-organization are not only useful at \textit{compressing} information about the properties of graphs which are changing dynamically, but that these can provide several insights about the dynamics itself. We should stress that these findings cannot be made general (i.e. to graphs and model other than the one studied), and they turn out being only descriptive, rather than \textit{predictive}. However, we believe that the graph entropy studied here does provide a compact way of obtaining information about the network to classify the dynamics. The entropy values considered here scale linearly in the number of nodes $N$, although in principle the number of dynamical variables scale quadratically in $N$. 
For graphs in which the number of edges is much larger than the number of nodes, this might allow for instance the use of \textit{heat maps} to classify and study the type of dynamics. 
%As a closing remark, we believe that this entropy has better asymptotic properties (long walks) as compared to those introduced in (\cite{shortpdv}), and thus can be used to study the thermalization properties of memristive circuits. 

In much more general terms, our study provides a connection between self-organization, non-locality and entropy-based measures. We thus hope our work will motivate further theoretical and experimental studies along these directions.

\section*{Acknowledgments}
We would like to thank Alioscia Hamma, Massimiliano Di Ventra and Fabio Traversa for discussion on some of the results of this work. 
We also acknowledge support from Invenia Labs.

\bibliographystyle{frontiersinHLTH&FPHY} % for Health and Physics articles
\bibliography{test}

%%% Upload the *bib file along with the *tex file and PDF on submission if the bibliography is not in the main *tex file

%%% Use this if adding the figures directly in the mansucript, if so, please remember to also upload the files when submitting your article
%%% There is no need for adding the file termination, as long as you indicate where the file is saved. In the examples below the files (logo1.jpg and logo2.eps) are in the Frontiers LaTeX folder
%%% If using *.tif files convert them to .jpg or .png

%\begin{figure}
%\begin{center}
%\includegraphics[width=8cm]{RandomKnockingPCP.pdf} \includegraphics[width=8cm]{RandomKnockingDP.pdf}
%\end{center}
%\textbf{\refstepcounter{figure}\label{fig:EPCP3} Figure \arabic{figure}.}{ Graph Entropy for various values of the particle creation probability as a function of time.}
%\end{figure}

%\begin{figure}
%\begin{center}
%\includegraphics[width=8cm]{Array.pdf}
%\end{center}
%\textbf{\refstepcounter{figure}\label{fig:ARRAY} Memristive circuit organized as a crossbar array..}
%\end{figure}

%%% If you don't add the figures in the LaTeX files, please upload them when submitting the article.

%%% Frontiers will add the figures at the end of the provisional pdf automatically %%%

%%% The use of LaTeX coding to draw Diagrams/Figures/Structures should be avoided. They should be external callouts including graphics.

\end{document}